\documentclass[a4paper,11pt]{article}
\usepackage{bbold}
\usepackage{tabu}

\usepackage{graphicx,subcaption}
\usepackage{amsfonts,amsthm,amsmath,amssymb,graphicx}
\numberwithin{equation}{section}
\usepackage{tensor}
\usepackage{yhmath}
\usepackage{epsfig}
\usepackage{slashed}
\usepackage{pdfpages}
\usepackage{graphicx,epstopdf}
\usepackage{hyperref}
\hypersetup{pdftex,colorlinks=true,allcolors=blue}
\usepackage{array}
\usepackage{tikz}
\usepackage{mathtools}
\usepackage{braket}
\usepackage{enumitem}
\usepackage{bm}
\usepackage{cite}
\usepackage[thinlines]{easytable}
\usepackage{cancel}

\newcommand{\ie}{{\it i.e.,}\ }
\newcommand{\bea}{\begin{eqnarray}}
\newcommand{\eea}{\end{eqnarray}}

\newcommand{\mO}{\mathcal{O}}

\newcommand{\mM}{\mathcal{M}}

\newcommand{\Op}{\mathcal{O}}

 \newcommand{\nn}{\nonumber}
\newcommand{\Tr}{\text{Tr}}

\newcommand{\eps}{\epsilon}
\newcommand{\sh}{\sinh}

\newcommand{\no}{\nonumber}
\newcommand{\der}{\partial}
\newcommand{\be}{\begin{equation}}
\newcommand{\ee}{\end{equation}}

\graphicspath{{figure/}}

\begin{document}

\title{Entanglement and geometry from subalgebras of the Virasoro algebra}

\author{Pawel Caputa and Dongsheng Ge}

\maketitle
\thispagestyle{empty}

\begin{center}
Faculty of Physics, University of Warsaw, ul. Pasteura 5, 02-093 Warsaw, Poland
\end{center}

\vskip 1cm

\begin{abstract}

  In this work we study families of generalised coherent states constructed from SL(2,R) subalgebras of the Virasoro algebra in two-dimensional conformal field theories. We derive the energy density and entanglement entropy and discuss their equivalence with analogous quantities computed in locally excited states. Moreover, we analyze their dual, holographic geometries and reproduce entanglement entropies from the Ryu-Takayanagi prescription. Finally, we outline possible applications of this universal class of states to operator growth and inhomogeneous quenches.

\end{abstract}

\clearpage
\pagenumbering{arabic} 

\section{Introduction}

One of the most important lessons for quantum gravity that we have learned from holography \cite{Maldacena:1997re} is the relation between the structure of quantum entanglement  and properties of gravitating spacetimes. Various studies starting from the celebrated Ryu-Takayanagi proposal \cite{Ryu:2006bv} and its generalisations \cite{Hubeny:2007xt,Nishioka:2009un,Rangamani:2016dms} even led to a  paradigm that the holographic geometry in Anti-de Sitter spaces (AdS) is in fact emergent from entanglement in dual conformal field theories (CFT) \cite{VanRaamsdonk:2010pw,Maldacena:2013xja}. This phenomenon is particularly manifest for the so-called thermofield double state that is holographically dual to the two-sided, eternal black hole \cite{Maldacena:2001kr}. The entanglement structure hidden in the  purification  is geometrically represented by the connectedness of the two black hole spacetimes that are otherwise  dual separately to two thermal density matrices. Further evidence has also been gathered in  more complicated holographic geometries that are dual to states in the so-called ``code sub-space" of holographic CFTs \cite{Almheiri:2014lwa}. 

Nevertheless, the number of CFT states  with clear and analytically tractable holographic dual geometries is still quite limited. The biggest progress has been achieved in two-dimensional CFTs where, despite the lack of an explicit example of a CFT dual only to  pure gravity\footnote{Interestingly, many properties of classical 3d gravity with the threading of massive particles can be recovered from an ensemble of CFTs with random CFT data \cite{Chandra:2022bqq}.}, the power of Virasoro symmetry can be harnessed to derive universal results valid for putative ``large-c" CFTs. On the other hand, holographically, by appropriately rendering the cut-off, a universal class of Bañados geometries \cite{Banados:1998gg} allows to construct metrics that reproduce these universal features of the 2d CFT states.

An interesting class of states that have a well-established gravity counterparts, that will be important in our work, consists of CFT states excited by local operators \cite{Nozaki:2014hna,Nozaki:2014uaa,Caputa:2014vaa,He:2014mwa,Caputa:2015tua,Chen:2015usa,Caputa:2014eta}. Their holographic dual involves a dynamical geometry that can be obtained from the  back-reaction of a massive particle (or generally a bulk field), which is  dual to that local operator \cite{Nozaki:2013wia}. In particular, entanglement evolution in these states has been studied extensively in the context of quantum quenches \cite{Asplund:2014coa,David:2022czg,Ageev:2022kpm,Kawamoto:2022etl,Caputa:2019avh,Bhattacharyya:2019ifi,Kusuki:2017upd,Kusuki:2019gjs}, scrambling \cite{Caputa:2015waa,Asplund:2015eha}, quantum chaos  \cite{Shenker:2013pqa,Roberts:2014ifa,Nie:2018dfe,David:2017eno} as well as bulk reconstruction in AdS/CFT \cite{Goto:2017olq,Kusuki:2019hcg,Kusuki:2019evw}. Certainly, this family provides very important and analytically tractable data points in the ``spacetime from entanglement" program.

In this work, we make some further progress in the above-mentioned program and consider an interesting general class of states in 2d CFTs excited by coherent action of higher Virasoro generators.  These states are among the generalized (Peremolov) coherent states \cite{Perelomov:book}. They have been recently utilized in various contexts such as for the study of the growth of operators and Krylov complexity in 2d CFTs \cite{Caputa:2021sib,Dymarsky:2021bjq} as well as exactly solvable deformations of CFTs \cite{Hikihara:2011mtb,Ishibashi:2016bey,Wen:2016inm,Tada:2019rls}.  Here, we will derive the expectation value of the energy momentum tensor and find a simple function that ``uniformizes" the answer into a Schwarzian derivative. From there, we obtain dual geometries that correspond to gravitationally dressed excitations in $AdS_3$ and have an interesting ``folded structure". We then compute a single-interval entanglement entropy and study its features in CFT as well as in gravity for different ranges of parameters.

This paper is organised as follows. To start with, in  section \ref{sec:VirCoh} we define our setup and coherent states. In section \ref{sec:SEexp} we compute the expectation value of the CFT stress tensor, uniformise it, and, in section \ref{sec:HolInt}, discuss its holographic interpretation. In section \ref{sec:EntEntropy} we discuss entanglement entropy in our coherent states. Last but not the least, in \ref{sec:SumDisc}, we discuss possible applications of our states and their dual geometries to Krylov complexity and inhomogeneous quenches, summarise and list some open problems. A few technical details are contained in three appendices.

\section{Virasoro coherent states}\label{sec:VirCoh}
We start by defining our setup and fixing conventions. Most of the arguments will be valid for general, two-dimensional CFTs with central charge $c$ (see e.g. \cite{di1997conformal,Ginsparg:1988ui} for standard reviews). The symmetries of such models are governed by two copies of infinite-dimensional Virasoro algebras with generators $L_m$ and $\bar{L}_m$ with $m\in \mathbb{Z}$ satisfying commutation relations
\begin{align}
[L_m, L_n] &= (m-n)L_{m+n} +\frac{c}{12}m(m^2-1)\delta_{m+n,0}\,,\label{VirasoroAlg}
\end{align}
and similarly for $\bar{L}_m$. We will refer to the copy of $L_m$ as chiral and $\bar{L}_{m}$ as anti-chiral. In the following, we will present our formulas keeping only the chiral part.

From this infinite set we pick a sub-set of three generators $\{L_{k},L_0,L_{-k}\}$ for some fixed positive integer $k$. From \eqref{VirasoroAlg}, we see that these generators close the following subalgebra
\be
[L_{0},L_{\pm k}]=\mp kL_{\pm k},\qquad [L_k,L_{-k}]=2kL_0+\frac{c}{12}k(k^2-1).\label{VirAl}
\ee 
In fact, by redefining the generators as
\be
 \tilde{L}_0=\frac{1}{k}\left(L_0+\frac{c}{24}(k^2-1)\right),\qquad \tilde{L}_{\pm 1}=\frac{1}{k}L_{\pm k},
\ee
we see that the triples $\{L_{k},L_0,L_{-k}\}$ form one of the infinitely many $SL(2,R)$ subalgebras of the Virasoro  \eqref{VirasoroAlg}.

Our main object of interest will be a family of generalised coherent states of Perelomov \cite{Perelomov:book} created by these Virasoro generators as follows\footnote{In general 2d CFT we will have two independent copies of the unitary operators acting on the highest weight state $\ket{h,\bar{h}}$}
\be\label{eq:viracoh}
\ket{\Psi_k(\xi)}=\exp\left(\xi L_{-k}-\bar{\xi}L_k\right)\ket{h},
\ee
where $\xi$ is a complex variable with complex conjugate $\bar{\xi}$ and $\ket{h}$ is the highest weight state such that
\be
L_0\ket{h}=h\ket{h}, \qquad L_{k}\ket{h}=0,\quad \text{for}\qquad k>0.
\ee
In \eqref{eq:viracoh}, the unitary operator acting on the highest weight state $\ket{h}$ is conventionally referred to as the displacement operator.\\
The motivation for considering these coherent states comes from various recent developments in high-energy as well as condensed matter studies of CFTs. For example, such states can be interpreted as ``universal" quantum circuits and studying their Nielsen's \cite{Chapman:2017rqy,Chagnet:2021uvi,Caputa:2018kdj,Erdmenger:2021wzc} or Krylov complexity \cite{Caputa:2021sib} is an active area of research. On the other hand, for purely imaginary $\xi=-it$, we can also view such states as quench evolution protocol (see e.g. review \cite{Calabrese:2016xau}) with a version of an inhomogeneous Hamiltonian of the SSD type that have been studied in \cite{Hikihara:2011mtb}. The goal of our work is to elaborate more on the interpretation of these states, including their holographic dual and analyze their entanglement structure.

For the purpose of performing computations, it will be useful to expand \eqref{eq:viracoh} in an orthonormal basis. For that we parametrize the complex coordinate as $\xi=\rho e^{i\theta}$ and apply the Baker–Campbell–Hausdorff formula (see appendix \ref{app:BCH}) to write
\be
\ket{\Psi_k(\rho,\theta)}=\frac{1}{\cosh^{2h_k}(k\rho)}\sum^{\infty}_{n=0}e^{in\theta}\tanh^n(k\rho)\frac{1}{n! k^n}L^n_{-k}\ket{h},
\ee
where we introduced\footnote{Notice that for $h=0$, $h_k$ becomes the dimension of the twist field
\be
h_k=\frac{c}{24}\left(k-\frac{1}{k}\right).
\ee
This resemblance of the ``orbifold" techniques will be present in various other steps of the analysis.}
\be
h_k=\frac{1}{k}\left(h+\frac{c}{24}(k^2-1)\right).
\ee
Orthonormal basis vectors are then defined as
\be
\ket{h,k,n}=\frac{1}{\sqrt{\mathcal{N}_{k,n}}}L^{n}_{-k}\ket{h},
\ee
were the normalisation is explicitly given by
\be
\mathcal{N}_{k,n}=\langle h|L^n_{k}L^n_{-k}|h\rangle=k^{2n}\prod^{n}_{l=1}l(2h_k+l-1)=n!k^{2n}\frac{\Gamma(2h_k+n)}{\Gamma(2h_k)}.
\ee
Finally, our three-parameter coherent states can be expanded in this basis as
\be
\ket{\Psi_k(\rho,\theta)}=\frac{1}{\cosh^{2h_k}(k\rho)}\sum^{\infty}_{n=0}e^{in\theta}\tanh^n(k\rho)\sqrt{\frac{\Gamma(2h_k+n)}{n!\Gamma(2h_k)}}\ket{h,k,n}.\label{CSkF}
\ee
Moreover, it will be useful to introduce a coordinate
\be\label{eq:zkdef}
z_k=e^{i\theta}\tanh\left(k\rho\right),
\ee
in terms of which the state is parametrized as
\be
\ket{\Psi_k(z_k,\bar{z}_k)}=(1-z_k\bar{z}_k)^{h_k}\sum^\infty_{n=0}z^n_k\sqrt{\frac{\Gamma(2h_k+n)}{n!\Gamma(2h_k)}}\ket{h,k,n}.\label{StateParamzs}
\ee
In the following sections we will explore these families of states labelled by different values of parameters $(k,\rho,\theta)$ and their entanglement structure. 
\section{Expectation value of the stress tensor}\label{sec:SEexp}
As a first step in this direction, we begin with computing the expectation value of the stress tensor. For 2d CFT on the plane, the stress tensor has only two independent components, chiral $T(z)$ and anti-chiral $\bar{T}(\bar{z})$. Again, we present the result for the chiral component. 

On the complex plane, the operator $T(z)$ is expanded in terms of the Virasoro generators \eqref{VirasoroAlg} as
\be
T(z) = \sum_{n\in \mathbb{Z}} z^{-n-2}L_n\,.
\ee
This way, suppressing the parameters of the state \eqref{StateParamzs}, we can first evaluate the following expectation value
\be
\langle\Psi_k| T(z)|\Psi_k\rangle =\sum_{n\in \mathbb{Z}} z^{-n-2}\langle\Psi_k| L_n|\Psi_k\rangle.
\ee
The details of this computation, even though they only involve the standard Virasoro algebra, are slightly involved and we included them in appendix \ref{app:Texp}. Here we only state the final result that is
\be\label{eq:SEexp}
\bra{\Psi_k}T(z)\ket{\Psi_k}=\frac{kh_kz^{2(k-1)}(1-z_k\bar{z}_k)^2}{(z^k-z_k)^2(1-z^k\bar{z}_k)^2}-\frac{c}{24}(k^2-1)\frac{1}{z^2}.
\ee
This formula allows us to ``geometrize" the coherent state in terms of a coordinate transformation. Namely, we can find a map $z\to f_k(z)$ such that
\be\label{eq:uniformeq}
\bra{\Psi_k}T(z)\ket{\Psi_k}
=\frac{c}{12}\{f_k(z),z\},
\ee
where $\{f(z),z\}$ is the Schwarzian derivative.\footnote{The Schwarzian is defined as \be
 \{ f(z),z\} = \frac{f'''(z)}{f'(z)} - \frac{3}{2} \left( \frac{f''(z)}{f'(z)} \right)^2 \,.\nonumber
\ee} Solving this ``uniformization" equation gives
\be
f_k(z)=\left(\frac{z^k-z_k}{z^k-\bar{z}^{-1}_k}\right)^{\alpha_k},\qquad \alpha_k=\sqrt{1-\frac{24h_k}{kc}}  = \frac{1}{k}\sqrt{1-\frac{24h}{c}}.\label{fkmap}
\ee
Note that the above function is uniquely determined up to a Möbius transformation that leaves the Schwarzian derivative intact. It is also interesting to compare our answer with the known result in the highest weight state
\be
\langle h|T(z)|h\rangle=\frac{h}{z^2}=\frac{c}{12}\{z^\alpha,z\},\qquad \alpha=\sqrt{1-\frac{24h}{c}}.
\ee
\subsection{Interpretation in terms of correlator with local operators}
Note that \eqref{fkmap} looks very similar to the function that uniformises one-point function of the stress tensor in states locally excited by primary operators. In fact we can check that our result is identical to the expectation value of the stress tensor with a local primary state 
\be \label{eq:Texpk}
\bra{\Psi_k}T(z)\ket{\Psi_k}=\frac{\langle \tilde{\Op}^\dagger(z_k)T(z^k)\tilde{\Op}(z_k)\rangle}{\langle \tilde{\Op}^\dagger(z_k)\tilde{\Op}(z_k)\rangle},
\ee
where the standard one-point function, fixed by conformal Ward identity\footnote{Can be computed using the OPE of $T(z)$ with the primary $\Op(z_1)$.}, reads
\be
\frac{\langle \Op^\dagger(z_1)T(z)\Op(z_1)\rangle}{\langle \Op^\dagger(z_1)\Op(z_1)\rangle}=\frac{\tilde{h}(1-z_1\bar{z}_1)^2}{(z-z_1)^2(1-z\bar{z}_1)^2}=\frac{c}{12}\{f_1(z),z\},
\ee
and the stress tensor is transformed under $z\to g(z)=z^k$ as
\be
T'(z)=g'(z)^2T(g(z))+\frac{c}{12}\{g(z),z\}=k^2 z^{2(k-1)}T(z^k)-\frac{c}{24}(k^2-1)\frac{1}{z^2}.
\ee
For this interpretation, and consistently with \eqref{fkmap}, we also have the conformal dimension $\tilde{h}$ of $\tilde{\Op}(z_k)$ given by
\be
\tilde{h}=\frac{1}{k}h_k.
\ee
This operator interpretation is not too surprising since the highest weight states in CFTs are usually associated with a mode of a primary operator 
\be
\Op(z)=\sum_{n\in \mathbb{Z}}z^{-n-h}\Op_n,
\ee
such that
\be
\ket{h}=\Op_{-h}\ket{0}=\lim_{z\to 0}\Op(z)\ket{0}.
\ee
This way, our states in CFTs can be thought of as simply
\be
\ket{\Psi_k}=U_k(z_k)\Op_h(0)\ket{0}.
\ee 
The action of this unitary displacement operator generally moves the operator so some position $z$. This is clear for $k=1$ where $L_{-1}$ is the momentum that indeed implies
\be\label{eq:globalcase}
e^{zL_{-1}}\Op(0)\ket{0}=e^{zL_{-1}}\Op(0)e^{-zL_{-1}}e^{zL_{-1}}\ket{0}=\Op(z)\ket{0}.
\ee
Nevertheless, for $k>1$ our finding \eqref{eq:Texpk} may be less familiar. 
In the following, given this intuitive picture and the bulk interpretation for local operator excitations  \cite{Nozaki:2013wia}, we will discuss the holographic counter-parts of our states also from the perspective of a massive particle connecting the two insertion positions of the operators at the boundary.
 
\section{Holographic Understanding}\label{sec:HolInt}
In this section, we consider holographic dual geometries of the coherent states discussed above\footnote{Strictly speaking in this part we have in mind the putative large-c 2d CFTs.}. As it is well-known, the Bañados ansatz \cite{Banados:1998gg} captures all the three-dimensional AdS spacetimes with flat asymptotic boundary, which will be our starting point in constructing the bulk dual of the Virasoro coherent states, given that we have obtained the stress energy tensors in the last section. In the following subsection \ref{sec:particleinter}, we give a more intuitive bulk understanding in terms of a massive particle moving from one boundary insertion point to the other. The back-reacted action can be evaluated through exercising a wedge from the pure AdS spacetime. The excised wedge introduces a logarithmic divergence into the action, which can be compared with the result  in \cite{Colin-Ellerin:2021jev,Hung:2011nu}. There they consider the bulk replica trick in the presence of a cosmic string (comparable to the massive particle in the dimension we are interested in). The metric they used for their calculation is the Skenderis-Solodukhin solutions \cite{Skenderis:1999nb} with delta-type singularities on the asymptotical boundaries. However, they claim the delta-type singularities are irrelevant once regulators are put properly around those singular points. Taking away the singularities brings their metric back to the Bañados type we are considering. This is to say
that the intuitive picture with the massive particle is effectively described by the Bañados geometry, which we rely on in the next section for the calculation of the holographic entanglement entropy. 

In the following part of this section, we first give an overview of the Euclidean Bañados geometries with emphasis on the correspondence to our Virasoro coherent states. Then, we present a detailed analysis on the massive particle interpretation for the $k=1$ coherent states.
Gauging the subtlety regarding the conical singularities considered in \cite{Colin-Ellerin:2021jev,Hung:2011nu}, we confirm that the Bañados approach and the massive particle picture give a consistent holographic understanding of the coherent states. 

\subsection{Bañados geometry and geodesic length}
In three dimensions, the most general solutions with flat asymptotic boundary are given by Euclidean Bañados metrics \cite{Banados:1998gg}
\be
ds^2=\frac{d\eta^2}{\eta^2}+\frac{(dz+\eta^2 \bar{\mathcal{L}}(\bar{z})d\bar{z})(d\bar{z}+\eta^2 \mathcal{L}(z)dz)}{\eta^2},\label{BanadosM}
\ee
where the holomorphic and anti-holomorphic functions parametrising the metric are
\be
\mathcal{L}(z)=\frac{3(f'')^2-2f'f'''}{4f'^2}=-\frac{1}{2}\{f(z),z\},\qquad \bar{\mathcal{L}}(\bar{z})=\frac{3(\bar{f}'')^2-2\bar{f}'\bar{f}'''}{4\bar{f}'^2}=-\frac{1}{2}\{\bar{f}'(\bar{z}),\bar{z}\}.
\ee
Locally, geometries \eqref{BanadosM} can be brought into Poincare coordinates
\be
ds^2=\frac{dZ^2+dw d\bar{w}}{Z^2},
\ee
with the explicit map for the boundary coordinates
\be
w=f(z)-\frac{2\eta^2(f')^2\bar{f}''}{4|f'|^2+\eta^2|f''|^2},\qquad \bar{w}=\bar{f}(\bar{z})-\frac{2\eta^2(\bar{f}')^2f''}{4|f'|^2+\eta^2|f''|^2},
\ee
as well as the change of radial coordinate
\be
Z=\frac{4\eta(f'\bar{f}')^{3/2}}{4|f'|^2+\eta^2|f''|^2}.
\ee
From the CFT perspective these maps can be interpreted in the language of ``uniformisation". More precisely, as soon as we manage to ``geometrize" a quantum state $\ket{\Phi}$ by finding coordinate transformations: $z\to f(z)$ and $\bar{z}\to \bar{f}(\bar{z})$, such that the expectation value of the stress tensor is
\be
\langle \Phi|T(z)|\Phi\rangle=\frac{c}{12}\{f(z),z\},\qquad \langle \Phi|\bar{T}(\bar{z})|\Phi\rangle=\frac{c}{12}\{\bar{f}(\bar{z}),\bar{z}\},
\ee
at large central charge, we immediately obtain dual gravity metrics\footnote{Generally, as will be also the case in our examples, we should be careful about singularities of the asymptotic metric. In that case, it is more appropriate to use solutions with curved asymptotic metric and we will be more careful about this below.} \eqref{BanadosM} with
\be
\mathcal{L}(z)=-\frac{6}{c}\langle \Phi|T(z)|\Phi\rangle,\qquad \bar{\mathcal{L}}(\bar{z})=-\frac{6}{c}\langle \Phi|\bar{T}(\bar{z})|\Phi\rangle.
\ee
One of the general results that can be derived from these local transformations is geodesic length. Indeed, having found $f(z)$ and $\bar{f}(\bar{z})$, we can compute the length of a geodesic $\gamma$ between two arbitrary points $(z_1,\bar{z}_1)$ and $(z_2,\bar{z}_2)$ at the boundary $\eta=\epsilon$ of the holographic geometry. The leading $\epsilon$ answer is 
\be\label{eq:geolength}
L_\gamma=\ln\left(\frac{(f(z_1)-f(z_2))^2}{f'(z_1)f'(z_2)\epsilon^2}\right)+\ln\left(\frac{(\bar{f}(\bar{z}_1)-\bar{f}(\bar{z}_2))^2}{\bar{f}'(\bar{z}_1)\bar{f}'(\bar{z}_2)\epsilon^2}\right).
\ee
This in turn can  be used to compute entanglement entropy or correlators of semi-classical operators from gravity. We will employ this in the next section while considering the  the entanglement entropy using the Ryu-Takayanagi prescription \cite{Ryu:2006bv}.


\subsection{Interpretation in terms of a particle in the bulk}\label{sec:particleinter}

\begin{figure}[h!]
  \centering
  \includegraphics[width=4cm]{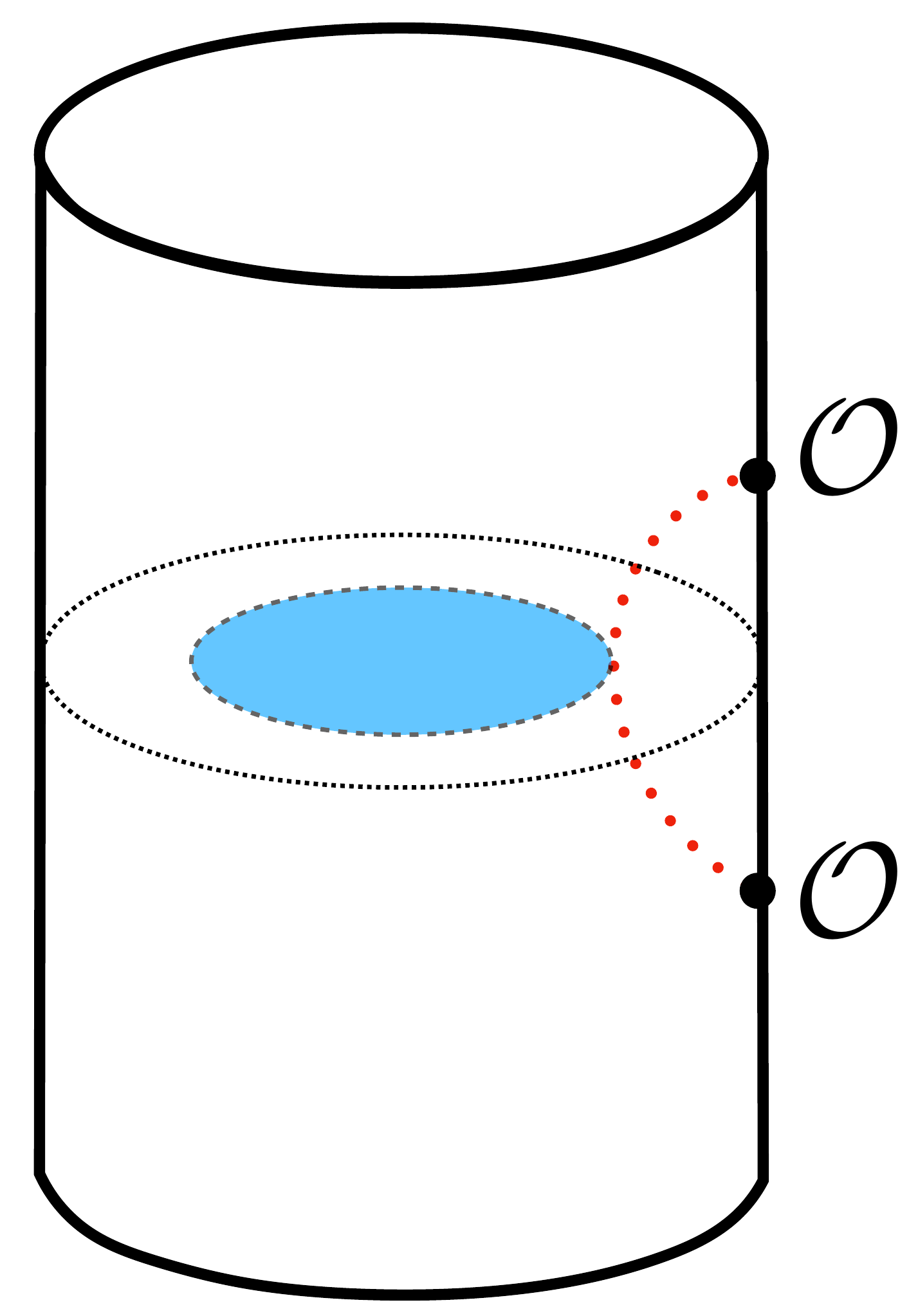}
  \caption{The massive particle moves along the red dashed geodesic which connects the two operators at the asymptotic boundary. The blue region can be interpreted as Krylov complexity (see discussions in section \ref{sec:SumDisc}).}
\label{fig:massiveparticle}
\end{figure}

  As shown in eq. \eqref{eq:globalcase}, the states built from the global conformal generators $L_{\pm 1}$ can be rewritten as a primary operator inserted in a shifted position with a certain normalization factor. This observation renders an interesting holographic interpretation for this class of states. Namely, they can be seen as a Euclidean AdS  with a massive particle ``following" a geodesic between the two insertion points of the primary operators  (see fig. \ref{fig:massiveparticle}) whose conformal weight are below the black hole threshold $h<\frac{c}{24}$ .

Below, we provide a detailed analysis of this setup.  We start by deriving the trajectory of the massive particle in Euclidean signature, then we consider the back-reacted geometry and evaluate the action on the excised geometry. 
As expected, the divergent part of the action is logarithmic, which comes from the volume of the excised region due to the backreaction of  the massive particle. This is consistent with the considerations in \cite{Colin-Ellerin:2021jev,Hung:2011nu} where authors evaluated the action using a more general Skenderis-Solodukhin solution \cite{Skenderis:1999nb} with delta-type singularities on the asymptotical boundary. 

Readers who are not concerned with  the technical details of this subsection can jump directly to the next section on the entanglement entropy.

\subsubsection{Trajectory of a massive particle}
The trajectory of the massive particle is  a timelike geodesic in the Lorentzian  AdS spacetime. However, here we are primarily interested in the Euclidean signature, it is no wonder that the geodesic can anchor on the two insertions of the primary operators. We work with  global coordinates $\{ \phi, \tau, \theta \}$ in the Euclidean signature and the metric is given by 
\be
ds^2 = \frac{l_{AdS}^2}{\cos^2 \phi} (d\tau^2 +d\phi^2 +\sin^2\phi d\theta^2)\,,
\ee
where $\phi = \pi/2$ is the AdS boundary. We take $l_{AdS}=1$ for simplicity in the following. A known stationary geodesic is in the coordinate center $\phi_s=0$, the subindex ``$s$'' is associated with the stationary one. We use the isometries of the spacetime to map the stationary one to that connecting the two primaries. This is done by applying a boost in the $(03)$ plane in the embedding coordinates (see eq. \eqref{eq:embeddingGL}), 
\begin{align}\label{eq:boost1}
&\cosh \tau_s \sec \phi_s =\sinh \beta  \cos (\theta-\theta_0)
   \tan \phi+\cosh \beta  \cosh \tau
   \sec \phi\,,\\
   &\sinh \tau_s\sec \phi_s =\sinh
   \tau \sec \phi\,,\\
  & \sin \theta_s  \tan
   \phi_s =\sin (\theta-\theta_0) \tan \phi\,,\\ \label{eq:boost4}
  & \cos
   \theta_s  \tan \phi_s =
\cosh
   \beta  \cos( \theta-\theta_0) \tan \phi+\sinh
   \beta  \cosh \tau \sec \phi\,,
   \end{align}
such that the stationary particle at the center ($\phi_s = 0$) is mapped to
\be\label{eq:geodesictau0}
\sin\phi = \frac{\cosh\tau}{\cosh\tau_0}\,,~~~ \theta=\theta_0\,,
\ee
 in the coordinate that we are working with. The boost angle is identified as $\cosh\tau_0 = -\coth \beta$ assuming $\tau_0>0$. To relate to the parameter of the coherent state $\ket{\Psi_1}$, the relation between $\tau_0$ and $\rho$ is $\tau_0 = \ln \coth \rho$. One feature of this geodesic is that it locates at a constant angular plane $\theta= \theta_0$ and anchors on the asymptotic boundary at $\tau = \pm \tau_0$. In principle, $\theta_0$ is undetermined through this coordinate transformation (the geodesic $\phi_s=0$ is symmetric under rotation). Another interesting feature  is that it is perpendicular to the asymptotic AdS boundary $\phi\to \pi/2$, as its  normal vector within the $\theta=\theta_0$ plane, $n_\mu \propto \cos\phi d\phi -\frac{\sinh\tau}{\cosh\tau_0} d \tau + 0 d\theta$, when evaluated on the asymptotic boundary, $n_\mu \sim- d\tau$. In the following part, without loss of generality, we take $\theta_0 = 0$.
     
\subsubsection{Action for the excised geometry}
The backreaction of the stationary particle 
 located at the center $\phi_s= 0$  can be understood in terms of exercising the AdS spacetime. One needs to  exercise a bulk wedge between $\theta_s = \eta $ and $\theta_s = -\eta~(0<\eta<\pi)$ , then identify the two surfaces together, as discussed in \cite{Matschull:1998rv,Balasubramanian:1999zv}.  The mass of the particle is related to the deficit angle of the exercised geometry  as
\be\label{eq:massdeficit}
m = \frac{2\eta}{\kappa}\,,
\ee
where $\kappa = 8\pi G$ is the gravitational constant.
For our setup, we need to map stationary geodesic to the one described in eq. \eqref{eq:geodesictau0} and exercise the corresponding wedge afterwards. Using \eqref{eq:boost1}-\eqref{eq:boost4},  the two surfaces $\theta_s = \pm\eta$,  are mapped to the surfaces 
\begin{align}\label{eq:surface}
&\Sigma_{\pm}:\pm\cot \eta \sin\theta=\sinh \beta  \cosh \tau 
   \csc \phi +\cosh \beta  \cos \theta \,,\\
  & \text{or}~~~\pm \cot \eta  \sin \theta =\cos \theta  \coth
   \tau_0-\cosh \tau 
   \text{csch}\tau_0 \csc \phi \,,
\end{align}
in our working coordinates. It is worth noting that the extrinsic curvature of the surfaces $\theta_s =\pm \eta$ vanishes, so as for that of  $\Sigma_{\pm}$.

\begin{figure}[h!]
  \centering
  \includegraphics[width=7cm]{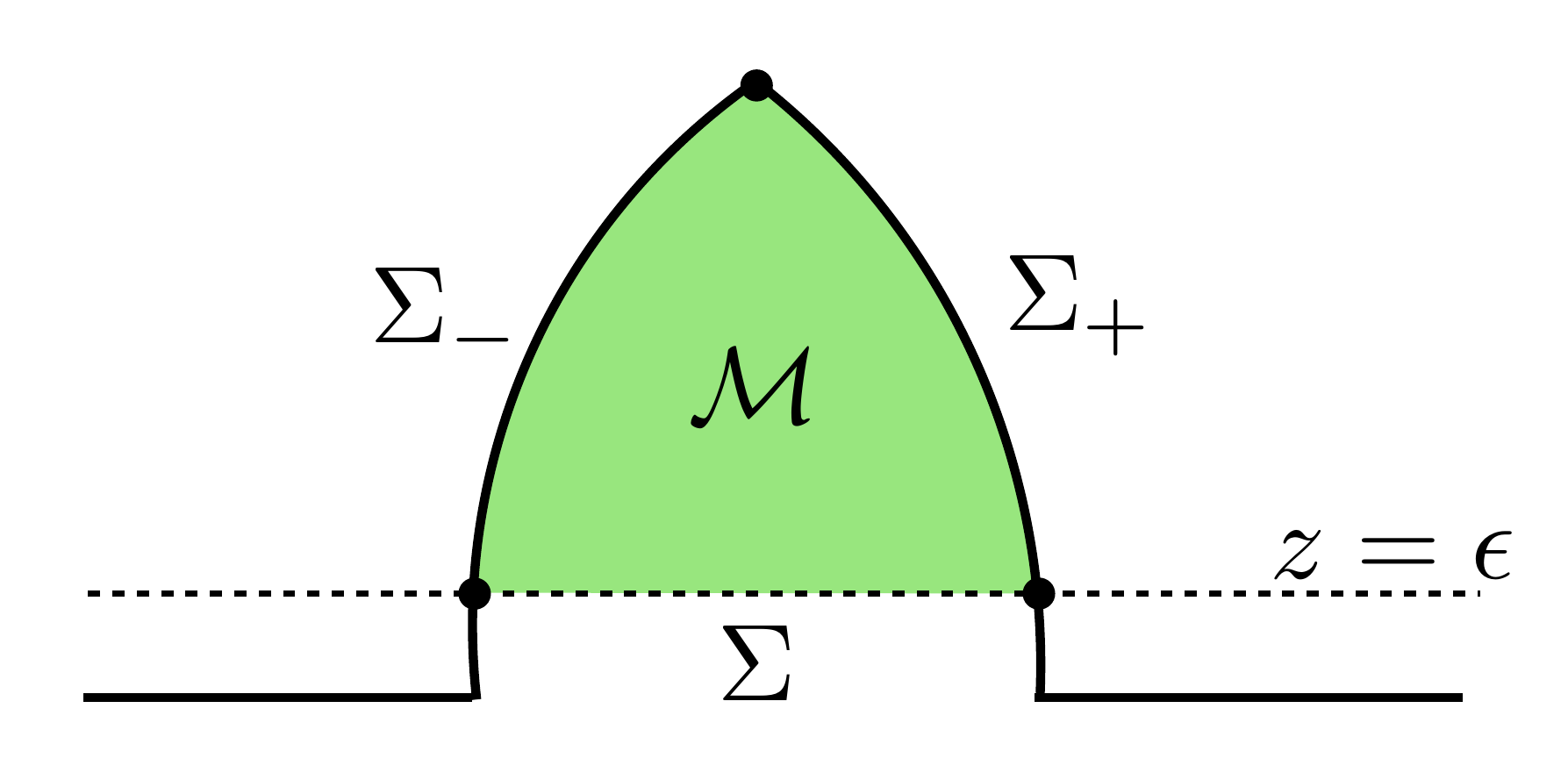}
  \caption{The slice is a constant time slice $t=0$ in Poincaré coordinates. $\mM$ is the wedge being exercised, which is bounded by three surfaces, the cutoff surface $\Sigma$ and $\Sigma_{\pm}$. }
\label{fig:ActionIll}
\end{figure}

\begin{figure}[h!]
  \centering
  \includegraphics[width=9cm]{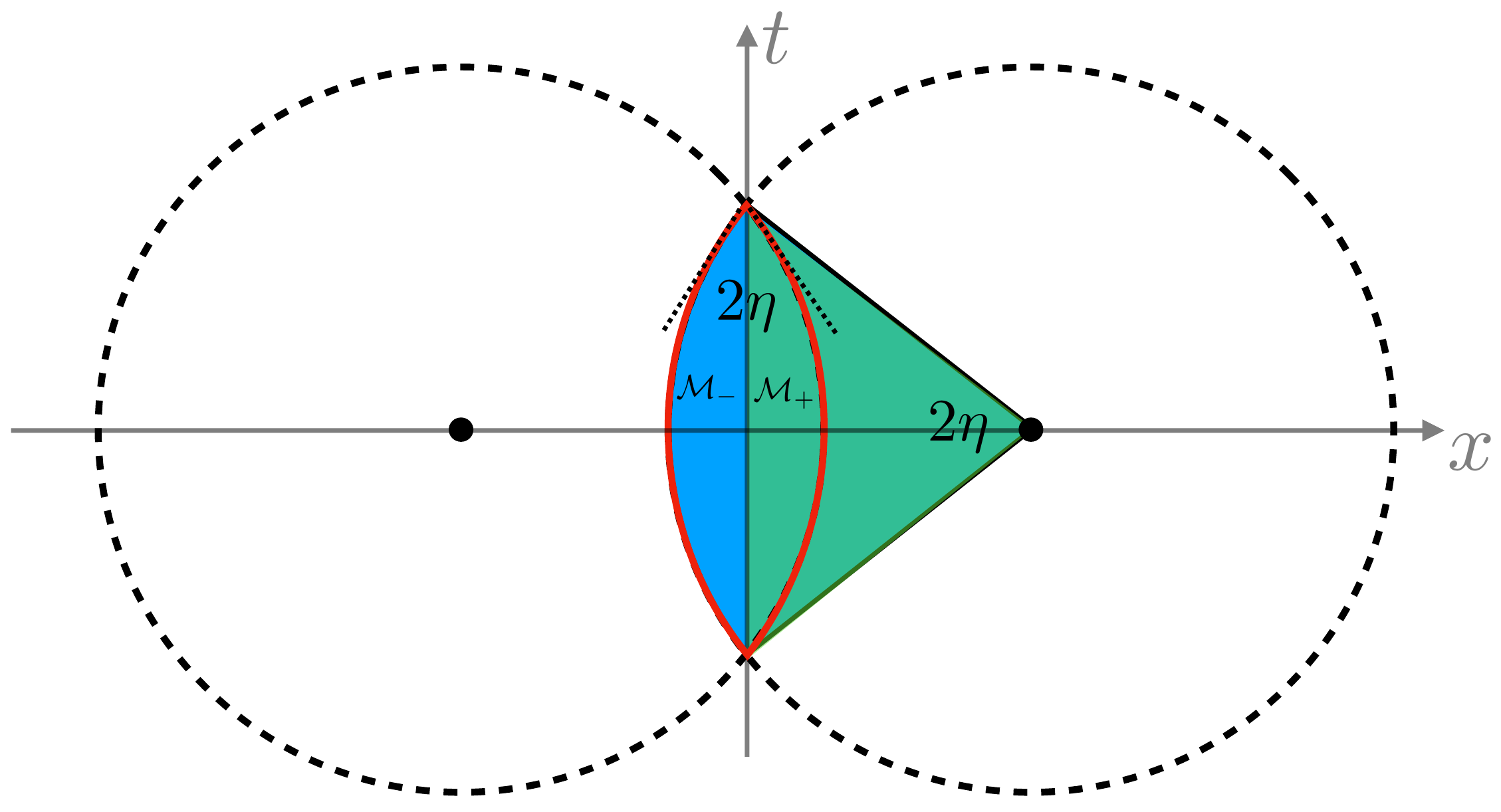}
  \caption{The projection of the excised wedge on the asymptotic boundary $z=0$ is the region within the red curve. The blue region is $\mM_-$, which is evaluated by subtracting the green region from the corresponding portion of the sphere. Its mirroring region within the red curve is $\mM_+$.}
\label{fig:BoundaryProjection}
\end{figure}

In the current situation, we are dealing with a simply connected Euclidean geometry without topological change after excising, there is no obstruction to the additivity of the gravitational action \cite{Brill:1994mb,Lehner:2016vdi}. We can evaluate the  action for the back-reacted geometry by subtracting the relevant action for the wedge $\mM$  and adding the action for the massive particle,
\be
I^{\text{BR}} = I^{\text{AdS}_3} - I^{\text{excised}} +I_m\,.
\ee
Upon the identification of $\Sigma_\pm$, a conical singularity arises along the trajectory of the massive particle. As pointed out in \cite{Chandra:2022bqq}, the contribution of the Ricci scalar on the conical singularity cancels the action of the massive particle.\footnote{We thank the anomalous referee for pointing this out to us.} 
The action for the AdS$_3$ part $I^{\text{AdS}_3} $ is UV finite once the counter term is introduced on the cutoff surface \cite{Balasubramanian:1999re}. Therefore, we only need to focus on the excised action $I^{\text{excised}}$ and find out the source for the divergence. It includes the following pieces
\begin{align}
 I^{\text{excised}} &= \frac{1}{2\kappa} I_{\mM}^{\text{EH}} 
 + 
 \frac{1}{\kappa}(I_{ \Sigma}^{\text{GH}} + I_{\Sigma}^{\text{ct}}) 
 +
    \frac{1}{\kappa} (I_{\Sigma \cap \Sigma_-}^{\text{Joint}}  +I_{\Sigma\cap \Sigma_+}^{\text{Joint}}  )\,, 
\\I_{\mM}^{\text{EH}}&= -\int d^3x \sqrt{g} (R + 2)\,,~~~I_{\Sigma}^{\text{GH}} = -\int_{\Sigma}d^2x \sqrt{h}K\,,~~~  I_{\Sigma}^{\text{ct}} =2\int_{\Sigma} \sqrt{h}\,,\no
\end{align}
where  $I_{\mM}^{\text{EH}}$ is the Einstein-Hilbert term for the excised region with constant Ricci curvature $R = -6$, $I_{\Sigma}^{\text{GH}}$ is the Gibbons-Hawking term on the surfaces surrounding the region $\mM$, as shown in fig. \ref{fig:ActionIll}, $I_{\Sigma}^{\text{ct}}$ is the counter term on the cutoff surface, while the remaining joint terms in the excised action are due to the non-smoothness on the surfaces upon identifying.  
Now we can evaluate the action piece by piece. It is simpler to do this calculation in the Poincar\'e coordinate $\{t,z,x\}$, which are related to the global coordinates via the following relations,
\begin{align}
z &= \frac{ \cos \phi}{\cos \theta \sin \phi+\cosh \tau}\,,~~~x = \frac{ \sin \theta  \sin \phi}{\cos \theta \sin \phi+\cosh
   \tau}\,,~~~t =  \frac{ \sinh \tau }{\cos \theta  \sin \phi +\cosh \tau }\,.
\end{align}
In this coordinate, the geodesic \eqref{eq:geodesictau0} satisfies 
\begin{align}
\frac{\left(t^2+x^2+z^2+l^2\right)^2}{4 \left(t^2+z^2\right)} = \coth^2 \tau_0\,,~~~\text{for}~~\theta_0\neq0\,,\\
z^2+t^2 =\tanh^2 (\tau_0/2)\,,~~~x=0\,,~~~\text{for}~~\theta_0=0\,,
\end{align}
where the geodesic equation in the first line reduces to the second one when taking $x=0$.
The surfaces $\Sigma_{\pm}$ in the Poincaré coordinates are mapped to 
\be
\Sigma_{\pm}: t^2 + z^2 + (x\pm \cot\eta \tanh(\tau_0/2))^2 =\csc^2\eta  \tanh^2(\tau_0/2)\,,
\ee
which are part of spheres centered at $t=z=0$ and $x = \mp \cot\eta \tanh(\tau_0/2)$ with radius $ \csc \eta  \tanh (\tau_0/2)$.
Let us choose the cutoff at $\Sigma:z= \eps$ and  start with the EH action for the part that $x<0$, which is half of the region $\mM = \mM_+ \cup \mM_-$,
\begin{align}
I_{\mM_-}^{\text{EH}} &= 4\int\frac{1}{z^3}dzdxdt \no\\
&= \frac{2}{\eps^2}\int dx dt + \int \frac{2\,dx dt}{t^2+(x-\cot\eta \tanh(\tau_0/2))^2 - \csc^2\eta  \tanh^2(\tau_0/2)}, 
\end{align}
the first term is divergent and will cancel with the Gibbons-Hawking term  $I_{ \Sigma}^{\text{GH}}$ and the counter term $I_{\Sigma}^{\text{ct}}$ on the cut-off surface $\Sigma$, which we will neglect from now on. Given that, we now focus on the evaluation of the second integral. 
It is better to work in the cylindrical coordinates,  
\be
x = \rho \cos\alpha + \cot\eta \tanh(\tau_0/2)\,,~~~ t = \rho \sin\alpha\,,
\ee
substituting inside the EH action part, the second term becomes
\begin{align}
\int \frac{2 \rho}{\rho^2 - \csc^2\eta  \tanh^2(\tau_0/2)} d\rho d\alpha \,,
\end{align}
for the integration region, we have to subtract the triangular region from the  disc sector  with opening angle $\delta=2\arccos\left( \frac{\cot\eta \tanh(\tau_0/2)}{\sqrt{\csc^2\eta   \tanh^2(\tau_0/2)-\eps^2}} \right)$ and radius $\rho=\sqrt{\csc^2\eta   \tanh^2(\tau_0/2)-\eps^2}$, as illustrated in fig. \ref{fig:BoundaryProjection}.

After careful calculation, neglecting the $O(\eps^{-2})$ term, the results are the following\footnote{For the integral on the triangle region, we have to use this indefinite integral when integrating over $\theta$,
\be
\int \ln \left( 1-\cos^2\eta \sec^2\alpha\right)d\alpha = \frac{i}{2}  \left(\text{Li}_2\left(e^{2 i (\alpha -\eta
   )}\right)+\text{Li}_2\left(e^{2 i (\eta +\alpha
   )}\right)-2 \text{Li}_2\left(-e^{2 i \alpha
   }\right)\right)\,.
\ee
}
\begin{align}
I_{\mM_-^1}^{\text{EH}} &=  -4\eta \ln \left( \frac{\tanh(\tau_0/2)}{\eps \sin\eta}  \right) +  O(\eps)\,,\\
I_{\mM_-^2}^{\text{EH}} & =  i \text{Li}_2\left(-e^{-2 i \eta }\right)+i \text{Li}_2\left(e^{2
   i \eta }\right) - \frac{i}{2} \text{Li}_2\left(e^{-4 i \eta }\right)
   +O(\eps^2)\,,
\end{align}
with $I_{\mM_-}^{\text{EH}} = O(\eps^{-2})+  I_{\mM_-^1}^{\text{EH}}-I_{\mM_-^2}^{\text{EH}}$, neglecting the leading divergence that will be cancelled by the GH term and the counter term. $\text{Li}_2(x)$ is the polylogarithmic function. The ``$+$'' region $I_{\mM_+}$ contributes equally as the ``$-$'' region, $I_{\mM_+}^{\text{EH}} = I_{\mM_-}^{\text{EH}}$.
Let us now consider the corner terms or Hayward terms \cite{Hayward:1993my}, the inner product between the out-pointing normals are
\be
n^{\Sigma} \cdot n^{\Sigma_-}=n^{\Sigma} \cdot n^{\Sigma_+} = -\eps \sin\eta\coth(\tau_0 /2)\,,
\ee
We have to keep in mind that upon identification, we should subtract an overall angle of $\pi$ from them. In a practical way, we subtract $\pi/2$ from each term, which gives
\begin{align}
I_{\Sigma \cap \Sigma_\pm}^{\text{Joint}} &= - \int \left( \cos^{-1}(n^{\Sigma} \cdot n^{\Sigma_\pm}) - \frac{\pi}{2}\right)  \frac{ \csc \eta  \tanh \left(\frac{\tau_0}{2}\right)}{\epsilon } d\theta
= 2\eta + O(\eps)\,.
\end{align}
Now we can  collect all the terms together due to excising, this gives
\begin{align}\label{eq:exercisedaction}
I^{\text{excised}} 
= &-\frac{4\eta}{\kappa} \ln \left( \frac{\tanh(\tau_0/2)}{\eps }  \right)  +\frac{4 \eta  }{\kappa }(1- \ln \sin\eta)\no\\
&-
\frac{2i \text{Li}_2\left(-e^{-2 i \eta }\right) + 2i \text{Li}_2\left(e^{2
   i \eta }\right) - i \text{Li}_2\left(e^{-4
   i \eta }\right) }{2\kappa} + O(\eps)\,,
   \end{align}
which  vanishes when $\eta\to 0$ as expected.
The cutoff $z=\eps$ in the Poincaré coordinates is related to the cutoff $\phi = \pi/2 -\tilde \eps$ in the global coordinates as $\eps = \frac{1}{2} \text{sech}^2{(\tau_0/2)}\tilde \eps+O(\tilde \eps^3)$. This allows us to  rewrite the first term in eq. \eqref{eq:exercisedaction}, \ie the logarithmic divergent term with  the $\tau_0$-dependent part, in terms of $\tilde \eps$, which gives
\be\label{eq:massaction}
-\frac{4\eta}{\kappa} \ln \left( \frac{2\sinh\tau_0}{\tilde\eps}\right) +O(\tilde\eps^2) =  -2m \ln \left( \frac{\coth\rho - \tanh\rho}{\tilde\eps} \right)+O(\tilde\eps^2) ,
\ee
where eq. \eqref{eq:massdeficit} and  the relation $\tau_0 = \ln \coth\rho$ are used
to relate to the parameter of the coherent state $\ket{\Psi_1}$.  For sufficiently small mass $m$ or small deficit angle $\eta$, one can identify the mass and the conformal weight of the scalar primary operator as $\Delta = h+\bar h =2h = m$.\footnote{The more precise relation between the scalar conformal weight and the massive particle is $\Delta = m(1-2Gm)$, for example in \cite{Chandra:2022bqq}, when the mass is small $m G\ll 1$, which gives  $\Delta \approx m$. We thank the anonymous  referee for raising concerns about this point.    } The logarithmic divergent part together with the $\tau_0$-dependent part is the same as in \cite{Colin-Ellerin:2021jev,Hung:2011nu}
upon identifying  $\tilde \eps = \delta$ and taking $\theta=0$ for the phase part of $u,~u'$. The divergent part of the back-reacted action can also be compared with the geodesic approximation of the propagator of the bulk scalar field as discussed \cite{Balasubramanian:1999zv}.

\section{Entanglement Entropy}\label{sec:EntEntropy}
For the coherent states \eqref{eq:viracoh}, the parameter $\rho$ determines the ``distance"\footnote{This can be made precise by considering e.g. the Fubini-Study metric in phase space.} from the the original highest-weight state. An interesting question then is how the entanglement entropy for one interval changes as $\rho$ grows. From the holographic side, the Ryu-Takayanagi formula computes this by the regulated geodesic length \eqref{eq:geolength}. Namely, for an interval $(z_1,\bar z_1)$ and $(z_2,\bar z_2)$,\footnote{For careful readers, $z_1$ and $z_2$ are simply the insertions of the operators, which have nothing to do with \eqref{eq:zkdef}. We will keep this notation for the rest of this section.} the holographic entanglement entropy of one interval is  given as
\be\label{eq:holEE}
S^{\text{Holo}}_k(z_1,z_2) = \frac{ L_\gamma}{4G}= \frac{c}{6} \ln \left| \frac{(f_k(z_1) - f_k(z_2))^2}{f'_k(z_1)f'_k(z_2) \eps_{UV}^2} \right|\,,
\ee
with functions $f_k(z)$ determined from solving the uniformization equation \eqref{eq:uniformeq}.

While in the CFT side,  we can use the replica trick to calculate the entanglement entropy \cite{Holzhey:1994we,Calabrese:2009qy}. Roughly speaking, the two-point function of the twist operators of conformal dimension $h_\sigma = \bar h_\sigma = \frac{c}{24}\left(n- \frac{1}{n} \right)$, inserted at the two ends of the interval computes the trace of the $n$-th power of the reduced density matrix $\rho^n_A$. Then the entanglement entropy is obtained taking the limit $n\to 1$,
 \be\label{eq:EE}
S_k^{\text{CFT}}(z_1,z_2) = -\lim_{n\to 1} \der_n \Tr \rho_{A,k}^n \,,~~~\Tr \rho_{A,k}^n \propto \bra{\Psi_k (\rho,\theta)}\sigma(z_1,\bar z_1) \tilde \sigma(z_2,\bar z_2) \ket{\Psi_k(\rho,\theta)} \,. 
\ee
For ``holographic" 2d CFTs, the above two-point function can be obtained using a coordinate transformation $f_k(z)$ which was found from the uniformization \eqref{fkmap}. Indeed using the transformation property of the primary operators we have
\begin{align}\label{eq:2ptcohbg}
 \bra{\Psi _k(\rho,\theta)}\sigma(z_1) \tilde \sigma(z_2) \ket{\Psi_k(\rho,\theta)}  &\sim (f'_k(z_1))^{h_\sigma} (f'_k(z_2))^{h_\sigma} \bra{0} \sigma(f_k(z_1))   \tilde \sigma(f_k(z_2)) \ket{0}\no\\
&= \frac{(f'_k(z_1) f'_k(z_2))^{h_\sigma}}{{(f_k(z_1) - f_k(z_2))}^{2h_\sigma}}\,.
\end{align}
 Up to a total constant which is a function of the UV cutoff, this way of calculation reproduces the holographic entanglement entropy \eqref{eq:holEE}. Moreover, in the  $k=1$ case, this two-point function is essentially a four-point function, of which the identity block dominates in the heavy-heavy-light-light (HHLL) scenarios\cite{Fitzpatrick:2014vua,Fitzpatrick:2015zha,Asplund:2014coa,Caputa:2014eta}. As we will show, closely following \cite{Asplund:2014coa}, the leading contribution of the HHLL block reproduces the holographic formula \eqref{eq:holEE}, thus consistently indicating the legitimacy of the uniformizing way of getting the two point function as in \eqref{eq:2ptcohbg}.  

Further, we utilize the HHLL block to analytically analyze the ``evolution"\footnote{We use evolution without reference to time, but to the parameter $\rho$ in the coherent state.} of the entanglement entropy as $\rho$ increases. For the more general $k\ge2$ case we just conduct the analysis by using the holographic formula or equivalently the uniformizing method.
\subsection{Entanglement Entropy from CFT and HHLL Block}
The coherent states constructed using global generators, 
as show in eq. \eqref{eq:globalcase}, may be written in terms of an operator inserted at a shifted location, which enables us to rewrite the two-point function of the twist operators in the coherent background to a four-point function
\be\label{eq:fourptresult}
 \bra{\Psi_1 (\rho,\theta)}\sigma(z_1,\bar z_1) \tilde \sigma(z_2,\bar z_2) \ket{\Psi_1(\rho,\theta)} = \frac{\langle \mO(u',\bar u') \sigma(z_1,\bar z_1) \tilde \sigma(z_2,\bar z_2) \mO(u,\bar u)  \rangle}{\langle \mO(u',\bar u')\mO(u,\bar u)  \rangle}\,,
\ee
where $u=e^{i\theta}\tanh\rho$ and $u'=e^{i\theta}\coth\rho$.\footnote{The notation here is again different from eq. \eqref{eq:zkdef}, but actually coincides with $k=1$ there.}
The conformal weight of the operator $\mO$ in our case is comparable to the central charge, $h/c \sim 1$, while the conformal weight of the twist operator in the limit $n\to 1$ is light compared to the large  $c\to \infty$. 
Given that we are considering a holographic CFT with a sparse spectrum of light operators, this enables us to use the expansion for the heavy-heavy-light-light block where the identity block dominates \cite{Fitzpatrick:2014vua,Asplund:2014coa}.  
With a conformal map $g(z) = \frac{(u'-z_1)(z-u)}{(u'-z)(z_1-u)}$ and $\bar g(\bar z) =\frac{(\bar u'-\bar z_1)(\bar z-\bar u)}{(\bar u'-\bar z)(\bar z_1-\bar u)} $, the insertions of the operators are mapped to $0,1,\infty$ and the cross ratio $\eta$, 
\be\label{eq:confmapping}
u'\to \infty\,,~~~  z_1 \to 1 \,,~~~ z_2\to \eta = \frac{(u'-z_1)(z_2-u)}{(u'-z_2)(z_1-u)}\,,~~~ u\to0\,,
\ee
analogously for the anti-holomorphic part, the expression \eqref{eq:fourptresult} becomes 

\be
\frac{\langle \mO(u',\bar u') \sigma(z_1,\bar z_1) \tilde \sigma(z_2,\bar z_2) \mO(u,\bar u)  \rangle}{\langle \mO(u',\bar u')\mO(u,\bar u)  \rangle}= G(\eta, \bar \eta) (1-\eta)^{2h_\sigma}(1-\bar \eta)^{2\bar h_\sigma}  \langle  \sigma(z_1,\bar z_1) \tilde \sigma(z_2,\bar z_2)\rangle\,,
\ee
where the leading contribution of the conformal block $G(\eta,\bar \eta)$ in its $t$-channel\footnote{In general, the four-point function doesn't have such a simple form, one needs to use e.g. recursion formula of Zamolodchikov  to do the calculation\cite{Zamolodchikov:1984eqp,1987TMP....73.1088Z}.} is
\be
G(\eta,\bar \eta ) =\left| \alpha ^{2h_\sigma} (1-\eta^\alpha)^{-2h_\sigma} \eta^{(\alpha-1)h_\sigma} \right|^2+ O\left( \left({h_\sigma}/{c}\right)^2 \right)\,.
\ee
We further substitute the above expression into eq. \eqref{eq:EE} for the entanglement entropy and introduce a UV scale $\eps_{UV}$ to keep the quantity dimensionless inside the logarithm, this gives
\be\label{eq:EECFT}
S(z_1,z_2) = \frac{c}{6}\ln \left| \frac{(1-\eta^\alpha)^2 \, \left( z_1-z_2\right)^{2}}{(1-\eta)^2 \alpha^2 \eta^{\alpha-1}\eps_{UV}^2}\right|\,.
\ee
 Obviously it  reproduces the holographic entanglement entropy \eqref{eq:holEE} once we substitute the uniformization solution \eqref{fkmap} into \eqref{eq:holEE}.
 
It is worthy to keep in mind that eq. \eqref{eq:EECFT} is the entanglement entropy for an interval on the complex plane. For the discussion below, we find it useful to subtract the vacuum contribution from eq. \eqref{eq:EECFT} such that its form is invariant under conformal transformations,
\be
\Delta S (\eta) =  S(z_1,z_2)  - S^{vac}(z_1,z_2) =  \frac{c}{6}\ln \left| \frac{(1-\eta^\alpha)^2 }{(1-\eta)^2 \alpha^2 \eta^{\alpha-1}}\right| = \frac{c}{3} \ln  \left| \frac{1}{\alpha} \frac{\sinh(\frac{\alpha \ln \eta}{2})}{\sinh{\frac{\ln \eta}{2}}} \right|\,,
\ee
which makes it more convenient to discuss the evolution on a cylinder through the map $z= e^{w} = e^{\tau + i \sigma}$ with $\tau$ being the Euclidean time and $\sigma$ the spacial direction.

\begin{figure}[h!]
  \centering
  \includegraphics[width=12cm]{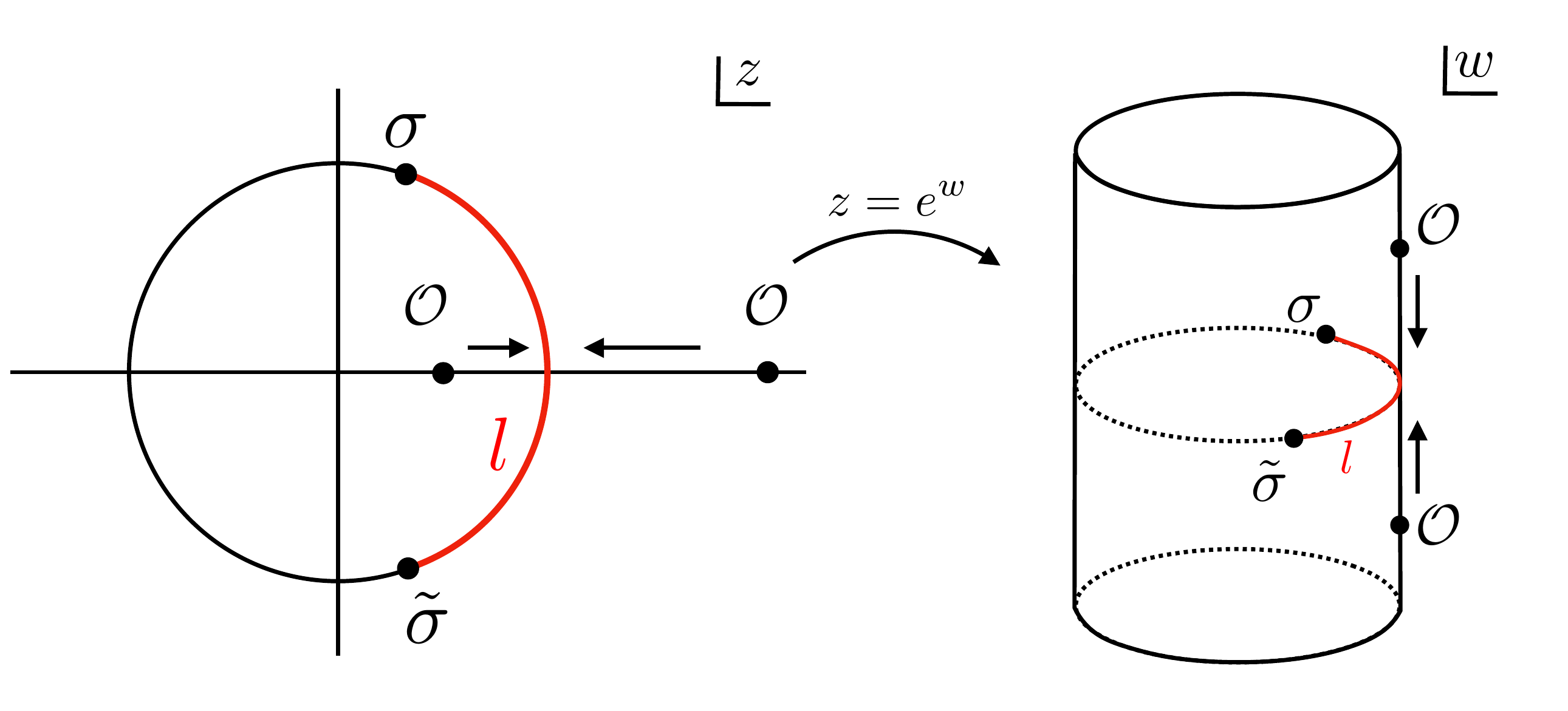}
  \caption{The setup for the evolution of the entanglement entropy on the complex plane (left) and on the cylinder (right). On the complex plane, the two heavy operators are placed on the real axis, while the twist operators are placed on a circle centered around the origin, symmetrically along the real axis; on the cylinder, the two heavy operators are placed vertically, while the twist operators are placed on a time slice, here we illustrate on $\tau =0$ slice.}
\label{fig:EEmoving}
\end{figure}

\subsection{Evolution of the Entanglement Entropy for $k=1$}
 Now we can use the result \eqref{eq:EECFT} to study the evolution of the entanglement entropy when the pair of heavy operators is approaching each other. 
Without loss of generality, we place the two heavy operators on the real axis of the complex plane with $\theta = 0$, and  the twist operators symmetrically around the real axis, as illustrated on the left side in fig. \ref{fig:EEmoving}
\be
u' = \coth\rho\,,~~~z_1 = re^{il/2}\,,~~~z_2 = re^{-il/2}\,,~~~ u =\tanh\rho\,.
\ee
It is clear that the coordinate depends on the radius of the circle $r$ (or different time slices on the cylinder), the angular difference of the two twist operators $l$ and the coherent state parameter $\rho$. Before considering the entanglement entropy, let us first take a look at the cross-ratio by substituting the above coordinates
\be
\eta=\frac{ (r e^{-\frac{i l}{2}}-\tanh \rho 
   ) (r e^{\frac{i l}{2}} -
   \coth\rho )}{(r e^{-\frac{i l}{2}}-\coth \rho ) (r e^{\frac{i l}{2}}   -  \tanh \rho )}\,.
\ee
It can be further decomposed into the real value and the imaginary one, given as
\begin{align}\label{eq:ReImeta}
&\text{Re}(\eta) =\frac{\frac{1}{4} \left(-2 \left(r^2+1\right) r \cos
   \left(\frac{l}{2}\right) \sinh (4 \rho )+r^2 \cos
   l (\cosh (4 \rho )+3)+\left(r^4+4 r^2+1\right)
   \sinh ^2(2 \rho )\right)}{AA^*}\,,\no\\
&\text{Im}(\eta) =- \frac{r \left(r \sin l \cosh (2 \rho )-\left(r^2+1\right) \sin \left(\frac{l}{2}\right)
   \sinh (2 \rho )\right)}{AA^*}\,,\no\\
   & (\text{Re}(\eta))^2 + (\text{Im}(\eta) )^2 = 1\,,~~~ A = \left(\sinh \rho -e^{\frac{i l}{2}} r \cosh \rho
   \right) \left(r \sinh \rho  - e^{\frac{i l}{2}} \cosh \rho \right)\,,
\end{align}
where we see that the cross-ratio $\eta$ always lies on the unit circle centered   around the origin and moves anticlockwise towards $\eta=1$ as $\rho$ increases to $\infty$. There are situations that the cross-ratio $\eta$ crosses the branch cut at $\eta =-1$, a necessary condition is that its imaginary part goes to zero, which means
\be\label{eq:crossingtime}
\cos (l/2)  = \frac{1+r^2}{2r} \tanh (2\rho)\,.
\ee
It is obvious that this equation can only be satisfied when $0<l\le\pi$, since $\rho\ge 0$. 

\begin{figure}[h!]
  \centering
  \includegraphics[width=4cm]{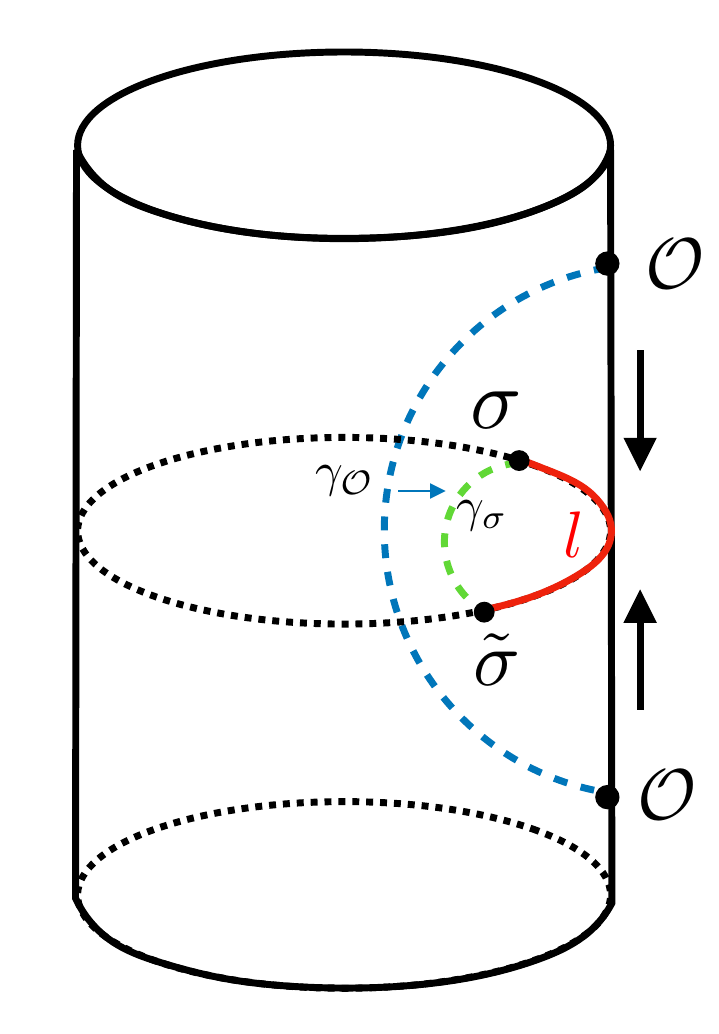}
  \caption{Bulk interpretation in terms of geodesic crossing. The blue dashed geodesic connecting the two operators $\mO$ is $\gamma_\mO$ while the green one connecting the two twist operators is $\gamma_\sigma$. When the two operators $\mO$ are moving towards each other, the blue geodesic shrinks, crossing the green one at the moment when the cross-ratio crosses the branch cut at $\eta=-1$. }
\label{fig:twogeos}
\end{figure}

In fact, once we map the setup to the cylinder (shown on the right side of fig. \ref{fig:EEmoving}), such a crossing of the branch cut can be understood holographically in terms of two geodesics, one is $\gamma_\mO$ connecting the two operators $\mO$ through the bulk  where the conical singularity locates, and the other is the geodesic  connecting the two twist operators $\gamma_\sigma$ placed on a given Euclidean time slice,\footnote{Different $r$ in \eqref{eq:crossingtime} correspond to different time slices on the cylinder. For the standard RT surface at $\tau=0$ we have $r=1$.} shown in fig. \ref{fig:twogeos}. 

\begin{figure}[h!]
  \centering
  \includegraphics[width=10cm]{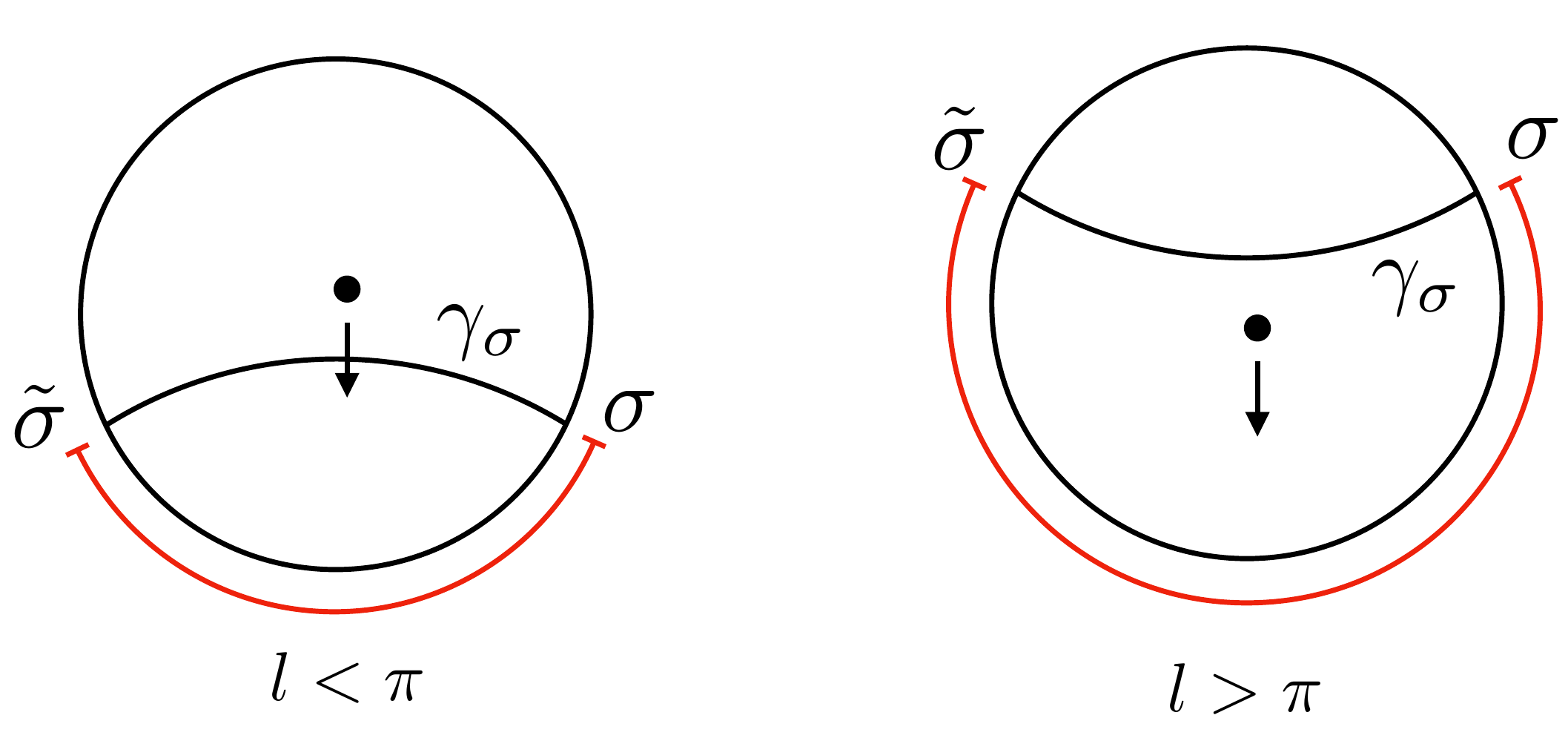}
  \caption{On a given time slice, the conical singularity is moving towards the AdS boundary. It intersects with the geodesic $\gamma_\sigma$ for the interval $l<\pi$ at a certain moment, while moving away from $\gamma_\sigma$ for $l>\pi$.}
\label{fig:holomoving}
\end{figure}

Initially when $\rho=0$, the conical singularity $\gamma_\mO$ is located at the coordinate center of AdS; as $\rho$ increases, $\gamma_\mO$ is moving towards the boundary and crosses $\gamma_\sigma$ at the moment when eq. \eqref{eq:crossingtime} is fulfilled; finally it shrinks  to a point on the boundary as $\rho\to \infty$. Clearly, if the interval $\pi<l<2\pi$, the two geodesics can never intersect with each other. We illustrate this in fig. \ref{fig:holomoving} on a given time slice.

\begin{figure}[h!]
  \centering
  \includegraphics[width=14cm]{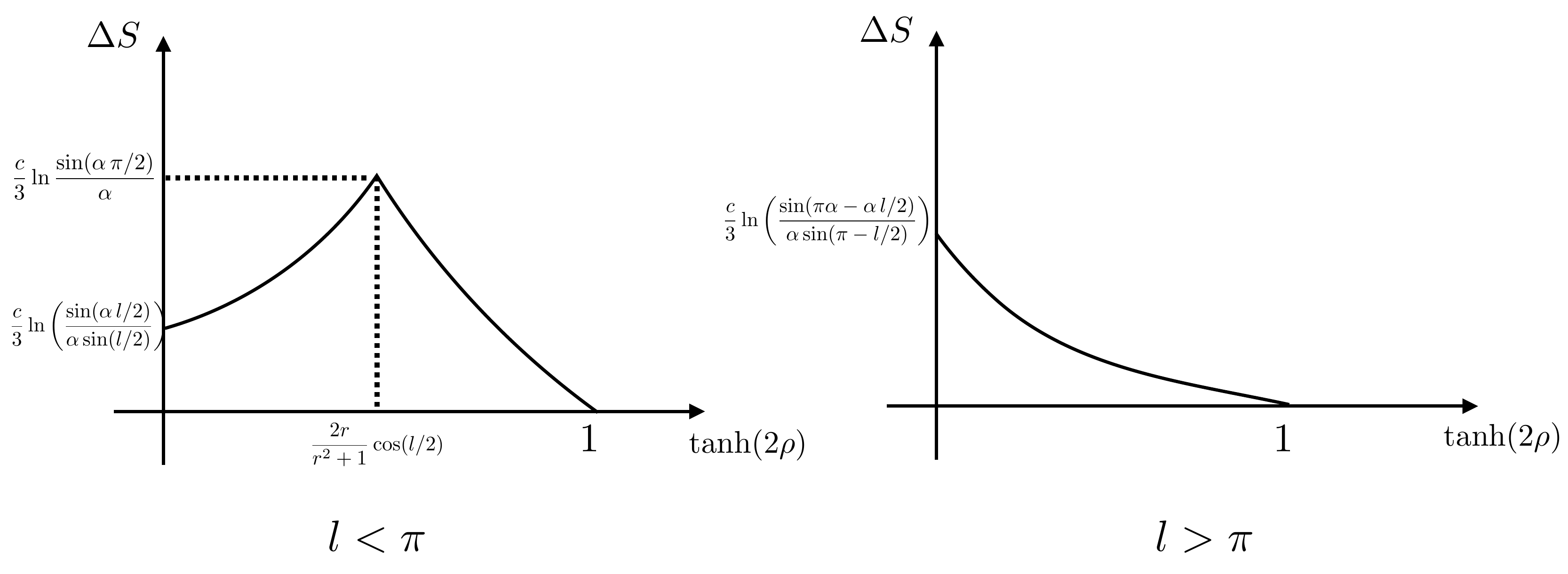}
  \caption{The evolution of the entanglement entropy after subtracting the vacuum contribution for the interval $l<\pi$ and $l>\pi$. The entanglement entropy difference $\Delta S$ asymptotes to zero as $\rho$ approaches infinity.}
\label{fig:EEresult}
\end{figure}

Now we move to the discussion of the evolution of the entanglement entropy as the two heavy operators are approaching each other. As we know, the cross ratio has a unit norm, such that it can be parametrized as $\eta = e^{i \sigma_\eta}$. However, the branches matter for the evaluation of the entanglement entropy. With the reparametrization in terms of the phase, the difference in the entanglement entropy is simply given by,
\be
\Delta S(\eta) = \frac{c}{3}\ln    \left| \frac{1}{\alpha} \frac{\sin(\frac{\alpha\,\sigma_\eta}{2})}{\sin{\frac{\sigma_\eta}{2}}} \right|\,,
\ee
which decreases monotonically with $\alpha$ and increases monotonically with $\sigma_\eta$ in the range $0<\alpha<1$ and $0<\sigma_\eta<\pi$. This simply tells us that $\Delta S(\eta)\ge0$ is non-negative.
To understand the evolution in more details, we need to distinguish the two cases, $l<\pi$ and $l>\pi$. For the case $l<\pi$, initially, $\sigma_\eta= -l$, combining with the vacuum entanglement entropy on the cylinder
\be
S^{cyl.\, vac.}(l) =\frac{c}{3}\ln \left(\frac{2 \sin (l/2)}{\eps_{UV}} \right)\,,
\ee
we obtain the known result when the conical singularity is at the center of the AdS
\be
S(l) = \frac{c}{3}\ln \left(\frac{2}{\alpha\, \eps_{UV}} \sin \left( \frac{\alpha\, l}{2} \right)\right)\,,
\ee
where as $\alpha $ becomes imaginary, this gives the entanglement entropy for the thermal state. As $\rho$ increases to $\frac{1}{2} \tanh^{-1}\left( \frac{2r}{r^2+1} \cos (l/2)\right)$, $\sigma_{\eta}$ decrease to $-\pi$, then jump to the second sheet and decrease to $\pi$ from $0$. As for the entanglement entropy difference, it increases initially  from $\frac{c}{3} \ln \left( \frac{\sin(\alpha\,l/2)}{\alpha \sin(l/2)}\right)$ to its maximal value
\be\label{eq:EEmaxk=1}
\Delta S_{\text{max}} = \frac{c}{3} \ln  \frac{\sin(\alpha\,\pi/2)}{\alpha}\,,
\ee
then decreases to $0$ in the end, as shown on the left side of fig. \ref{fig:EEresult}.

For $l>\pi$, there is no crossing, the entanglement entropy is evaluated directly in the second sheet. Initially $\sigma_\eta = 2\pi - l$, as $\rho$ increases, $\sigma_\eta$ decreases to $0$. The entanglement difference decreases monotonically to $0$ from $\frac{c}{3} \ln \left( \frac{\sin(\pi\alpha-\alpha\,l/2)}{\alpha \sin(\pi-l/2)}\right)$, as shown on the right side of fig. \ref{fig:EEresult}.

In both cases, the entanglement entropy is always larger than the vacuum entanglement entropy and decreases to the level of the vacuum entanglement entropy in the end. This large $\rho$ behavior is consistent with the result in \cite{Asplund:2014coa} for considering local quenches inside the interval at early stages. This can also be understood in terms of the energy density, when $\rho\to \infty$,
\be
T(z) \stackrel{\rho\to \infty}{\sim} \frac{h}{e^{4\rho} (z-1)^4}\sim 0\,,
\ee
if $\rho$ approaches to the infinity faster than $z$ approaches $1$, which is essentially the vacuum energy up to contact terms. Therefore, it is of no surprise that we obtain the vacuum entanglement entropy in the end.

\subsection{Evolution of the Entanglement Entropy for general $k\ge2$}

For the general case $k\ge2$, we can use the uniformization method or the holographic way to study the evolution of the entanglement entropy as one of the parameters grow. To start with, we introduce  the $k$-folded coordinate $w=z^k$ and  define a function analogous to the cross-ratio $\eta$ in the $k=1$ case\footnote{The notation here is different from eq. \eqref{eq:zkdef}, but $u_k=z_k$ in that notation. }
\be\label{eq:crossratiok}
\eta_k = \frac{w_2 - u_k}{w_2 - \bar u_k^{-1}}\frac{w_1 - \bar u_k^{-1}}{w_1 - u_k}\,,\qquad
~~~u_k=e^{i\theta}\tanh\left(k\rho\right)\,.
\ee
Substituting the uniformizing function \eqref{fkmap} to the entanglement entropy formula allows us to write
\begin{align}
S_k(z_1,z_2) &= \frac{c}{6}\ln \left| \frac{(f_k(w_1) - f_k(w_2))^2}{f_k'(w_1) f_k'(w_2)\eps_{UV}^2} \left( \frac{\der w_1}{\der z_1} \right)^{-1} \left( \frac{\der w_2}{\der z_2} \right)^{-1} \right|,
\no\\
&=\frac{c}{6}\ln \left| \frac{(1-\eta_k^{\alpha_k})^2 \,(w_1-w_2)^2 }{(1-\eta_k)^2 \alpha_k^2 \eta_k^{\alpha_k-1}\eps_{UV}^2} 
\frac{1}{k^2 z_1^{k-1} z_2^{k-1} }
\right|\,.
\end{align}
From the expression above, the first factor inside the log has a similar form as in eq. \eqref{eq:EECFT}, while the second factor is  a result of the chain rule. From another point of view, regarding $w=z^k$ as a conformal map, the second factor essentially comes from the conformal factor for the twist operators during the conformal transformation. The entanglement entropy in the folded coordinate is then given by
\be\label{eq:kfoldEE}
S_k(w_1,w_2) = \frac{c}{6}\ln \left| \frac{(1-\eta_k^{\alpha_k})^2 \,(w_1-w_2)^2 }{(1-\eta_k)^2 \alpha_k^2 \eta_k^{\alpha_k-1}\eps_{UV}^2} 
\right|
\ee
which resembles \eqref{eq:EECFT} in the $k=1$ case with the new ``cross-ratio'' defined as \eqref{eq:crossratiok} and $\alpha_k = \alpha/k$. This is consistent as we have noticed for the expectation value of the stress energy tensor \eqref{eq:Texpk}.

\begin{figure}[h!]
  \centering
  \includegraphics[width=14cm]{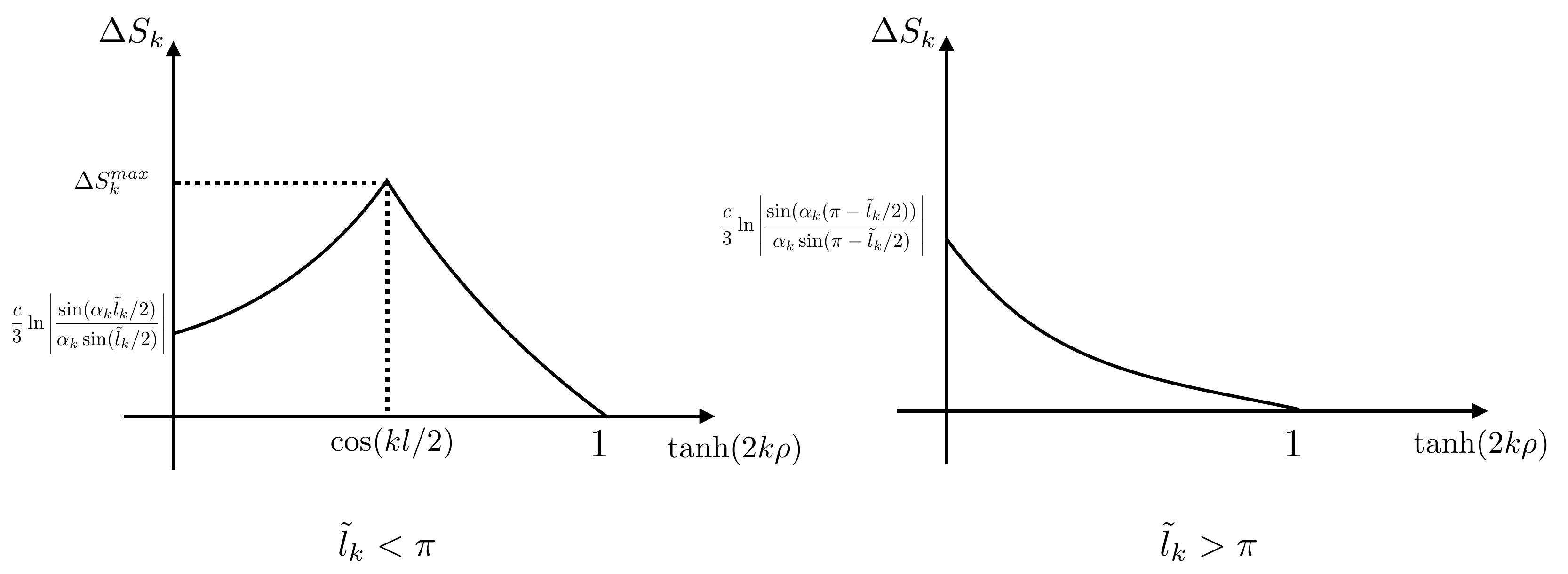}
  \caption{The evolution of the entanglement entropy after subtracting the ``k-folded vacuum'' contribution for the interval $\tilde l_k<\pi$ and $\tilde l>\pi$. }
\label{fig:EEevok}
\end{figure}

We proceed by placing $z_1 = e^{il/2}$, $z_2 = e^{-il/2}$ and taking $\theta=0$ for the coherent state as in the previous subsection, which means that in the folded patch $w_1 = e^{ilk/2}$ and $w_2 = e^{-ilk/2}$. Subtracting from eq. \eqref{eq:kfoldEE} the ``$k$-folded vacuum''\footnote{By  ``$k$-folded vacuum'', we mean that the two-point correlator of the twist operators evaluated in such a state is $\langle \sigma(w_1)\sigma(w_2) \rangle_{\text{$k$-fold}} = (w_1- w_2)^{-2 h_\sigma}$. } contribution
gives
\begin{align}\label{eq:EEdiffk}
\Delta S_k(e^{ilk/2},e^{-ilk/2}) &= \frac{c}{6}\ln \left| \frac{(1-\eta_k^{\alpha_k})^2 \, }{(1-\eta_k)^2 \alpha_k^2 \eta_k^{\alpha_k-1}}\right| ,
\\
&=\frac{c}{3}\ln \left| \frac{\sh\left(\frac{\alpha_k}{2}\ln\eta_k \right)}{\alpha_k \sh \left( \frac{1}{2}\ln\eta_k\right) }\right|,
\end{align}
where  $\ln \eta_k$ returns the principle value for the following analysis and  more explicitly $\eta_k$ is given by
\be
\eta_k = \frac{e^{-ilk/2} -\tanh(k \rho)}{e^{-ilk/2} -\coth (k \rho)} \frac{e^{ilk/2}-\coth(k\rho)}{e^{ilk/2} -\tanh(k \rho)}\,,
\ee
 Similarly as in \eqref{eq:ReImeta}, we can perform an analysis for the entropy difference in terms of the ``cross-ratio''. Not surprisingly, $|\eta_k| = 1$ lies on the unit circle as well. For $\rho =0$, $\eta_k = e^{-i\tilde l_k}$ with $\tilde l_k \equiv lk \mod 2\pi$; as $\rho$ increases,\footnote{Here $k$ is fixed. Alternatively, one can fix $\rho$ and increase $k$, the effect will be the same.} the phase of $\eta_k$ decreases and reaches to zero $(\text{mod}~ 2\pi)$ as $\rho$ goes to infinity. Depending on the initial phase, the evolution of the entanglement entropy exhibits two different behaviors. In the case $\tilde l_k <\pi$, the entropy difference \eqref{eq:EEdiffk} grows to its maximum  when
\be\label{eq:EEmaxgk}
\cos(kl/2) = \tanh(2k\rho)\,,~~~\text{with} ~~~\Delta S_k ^{max} = \frac{c}{3} \ln\left| \frac{\sin(\alpha_k \pi/2)}{\alpha_k} \right|\,,
\ee
before dropping down to zero. This condition can also be interpreted as the crossing of two geodesics, however, in the $k$-folded patch, \ie in the $w$-coordinate. For $\tilde l_k> \pi$, the entropy difference decreases monotonically to zero
\be
\lim_{\rho\to \infty} \Delta S_k(e^{ilk/2},e^{-ikl/2}) = 0 \,.
\ee
The two scenarios are plotted in fig. \ref{fig:EEevok}, which resembles the behavior in the $k=1$ case.

Note that the results look identical to the $k=1$ when working with effective parameters $\tilde{l}_k=lk \mod 2\pi$, $\alpha_k = \alpha/k$ and $\rho_k=k\rho$. However, one should keep in mind that the physical length of the entangling region is still $l$ instead of $\tilde l_k$ i.e., one of the effects of our coherent state of $L_k$'s is the effective rescaling of degrees of freedom in the entangling region from $l\to k l$. Besides that,  a rescale of $k$ makes it faster for the entanglement entropy difference to approach zero asymptotically as $\rho$ grows to infinity. 

\section{Summary and Discussions}\label{sec:SumDisc}
We conclude this work with some discussions and suggestions for future studies. The whole study circles around different classes of generalized coherent states, constructed with the $SL_k(2,\mathbb{R})$ sub-sectors of the Virasoro generators acting on the highest weight state  \eqref{eq:viracoh}. We obtain the energy density distributions through explicit calculations of the expectation values of the stress energy tensor \eqref{eq:SEexp} using CFT techniques. This enables us to gain a holographic understanding of these coherent states in terms of  the Bañados solution. More specifically for the class of the coherent states constructed using global conformal generators $L_{\pm 1}$ and $\bar L_{\pm 1}$  when $k=1$, we can interpret the bulk picture  in terms of a massive particle moving along a certain Euclidean geodesic which anchors asymptotically on the insertion points of the operators, as in fig. \ref{fig:massiveparticle}. This is based on the properties of the global conformal generators so as to render an operator interpretation of this class of coherent states. Analogously, the general $k>1$ cases can have a particle interpretation as well, though in the $k$-folded patch. However, this should be treated with cautions as the insertion of the operators in the $k$-folded picture is merely an effective description.

As an interesting application, we study the entanglement property of such states, with emphasis on the  evolution of one-interval entanglement along the increment of the norm of the complex number $\xi$ parametrizing the coherent states. Holographically, this is calculated in terms of the regulated length of the geodesic which anchors asymptotically on the two ends of the interval using the RT formula \cite{Hubeny:2007xt,Roberts:2012aq,Sheikh-Jabbari:2016znt}, which matches the CFT calculations using the replica trick and uniformization method. Especially when $k=1$, the two-point correlation function of the twist operators can be rewritten as a four-point correlator of HHLL type, where we show that the leading contributions from the HHLL conformal block is equivalent to the two-point function using the uniformization method, as in eq. \eqref{eq:EECFT}. We further study the evolution of  entanglement entropy of a single interval $l$ on the cylinder in both the $k=1$ and the general $k>1$ family of excited states. For all the cases, there are two scenarios depending on the interval size, which are plotted in fig. \ref{fig:EEresult} and fig. \ref{fig:EEevok}. When the interval size $l<\pi$ or the effective one $\tilde l_k<\pi$ (though the physical size of the entangling region is still $l$), there exists a peak for the entanglement difference, given in eq. \eqref{eq:EEmaxk=1} and \eqref{eq:EEmaxgk}, which can be interpreted in terms of geodesic crossing in the AdS or the $k$-folded AdS space. While for the interval size $l>\pi$ or the effective one $\tilde l_k>\pi$, the entanglement difference decreases monotonically to zero. Pictorially, this can be understood via the motion of two operators with the level $k$ indicating the relative velocity. Along the increment of $|\xi|$, the two operators approach each other and finally fuse together to an identity block, which explains the vanishing of the difference in the entanglement entropy.

Certainly, the setup we have considered is not  confined to the study for the entanglement entropy. Rather, it can host quite a wide range of applications in the framework of 2d CFTs, among which we will discuss two in the following part.

 The first one will be the operator growth and the so-called Krylov complexity \cite{Parker:2018yvk}.
In that context, one can consider the growth of some abstract operator governed by the Liouvillian super-operator that, in the Krylov basis, can be represented by the ladder operators of some symmetry algebra \cite{Caputa:2021sib} (see also \cite{Patramanis:2021lkx}). For example, we may be interested in the Liouvillian of the type
\be
\mathcal{L}_{(k)}=\alpha(L_{-k}+L_k),
\ee
such that the state representing the operator growth becomes
\be
\left|\Op(t)\right)=e^{i\mathcal{L}_kt}\ket{h}=\sum^\infty_{n=0}\frac{\tanh^n(k\alpha t)}{\cosh^{2h_k}(k\alpha t)}\sqrt{\frac{\Gamma(2h_k+n)}{n!\Gamma(2h_k)}}\ket{K_n}\equiv \sum^\infty_{n=0}\phi_n(t)\ket{K_n},
\ee
where $\ket{h}$ is the highest weight state and $\ket{K_n}$ stand for the Krylov basis vectors.\\
This is precisely our state \eqref{CSkF} with identification
$\rho=\alpha t$ and $\theta=\pi/2$. Following \cite{Caputa:2021sib}, this can be interpreted as a motion in phase space along this trajectory (geodesic in Fubini-Study metric space).\\
Moreover, the growth of the operator in this protocol can be characterised by the Krylov complexity\footnote{See \cite{Parker:2018yvk,Dymarsky:2021bjq,Kar:2021nbm,Caputa:2021sib,Balasubramanian:2022tpr,Rabinovici:2022beu,Muck:2022xfc,Bhattacharjee:2022vlt,Hornedal:2022pkc,Balasubramanian:2022dnj} for some recent developments pedagogical reviews.} 
\be
\mathcal{K}_\Op=\sum_n n|\phi_n(t)|^2=2h_k\sinh^2(k\alpha t).
\ee
One can actually show that this quantity can be expressed by the variation of the $SL(2,R)$ symmetry generator $L_0$. Namely we have the relation \cite{Caputa:2021ori}
\be
\mathcal{K}_\Op=(\Op(t)|L_0|\Op(t))-(\Op(0)|L_0|\Op(0)).
\ee
This variation of the expectation value of $L_0$ can be easily computed using our dual Bañados metrics and their associated expectation value of the holographic stress tensors. Indeed from the general expression for expectation values in some state $\ket{\Phi}$, we have
\be
\langle \Phi| L_0|\Phi\rangle=\frac{1}{2\pi i}\oint dz\, z\, \langle \Phi| T(z)|\Phi\rangle=-\frac{c}{12\pi i}\oint dz\, z\, \mathcal{L}(z).
\ee
In addition, another geometric relation between $\mathcal{K}_\Op$ and volume on the hyperbolic disc was observed in \cite{Caputa:2021sib}. It is tempting to identify the Fubini-Study hyperbolic disc (i.e., the information metric \cite{Miyaji:2015woj}) with the $\tau=0$ slice of the Bañados metric dual to our coherent state. The geodesic of the massive particle that we discussed crosses this slice at point $\rho=\alpha t$ and the volume (area) of the disc (the blue shaded region in fig. \ref{fig:massiveparticle}) from the origin of AdS up to that point is proportional to $\mathcal{K}_\Op$. Though at the moment the above statements are merely mathematical observations, it is worth to explore these heuristic interpretations further and test whether they can be elevated to the proper holographic dual of the Krylov complexity.
Another new direction in the Krylov complexity is the interpretation of the universal operator evolution in 2d CFTs in terms of the Young lattice \cite{Caputa:2021ori} (see also \cite{Besken:2019jyw}). In that setup, operator $\mathcal{L}_{(k)}$ adds or subtracts $k$ boxes to and from Young diagrams, so for $k>1$ becomes non-local.\footnote{The local $k=1$ case was analyzed in \cite{Caputa:2021ori}.} It will be interesting to understand this non-local growth and fastest/slowest paths through the lattice, also from the matrix model perspective (along the lines of \cite{Chattopadhyay:2019pkl}).

The second interesting application of our coherent states and the associated dual geometries might be in the context of the so-called inhomogeneous 2d CFTs, whose Hamiltonian is obtained through a convolution of the (undeformed) Hamiltonian density $h(x)$ and an inhomogeneity function $f(x)$, 
\be
H(f)=\int dx\,f(x)h(x).
\ee
From the geometric perspective, such a deformation can be equivalently seen as placing the 2d CFT on a curved background whose time component becomes space-dependent
\be
ds^2=f(x)^2d\tau^2+dx^2.
\ee
For instance, considering a function on the cylinder ($x\sim x+ 2\pi$)
\be
f(x)=\gamma+2\alpha\cos(kx), 
\ee
the corresponding Hamiltonian for the inhomogenous CFT would be
\be
H(f)=\gamma L_0+ \alpha(L_k+L_{-k})+\text{anti-chiral}.
\ee
In fact, this type of Hamiltonian has appeared in the study of quantum quenches, such as the Möbius quench and the SSD quench.\footnote{See also \cite{Czech:2019vih,deBoer:2021zlm,Khetrapal:2022dzy,Lapierre:2020ftq,Lapierre:2019rwj,Moosavi:2019fas,Langmann:2018skr,Gawedzki:2017woc} for similar operators playing the role of the modular Hamiltonian in 2d CFTs.}
The techniques  and the dual bulk geometries in our work can serve as building blocks for the holographic understanding of the quench dynamics in those deformed models or related questions with excited states. Some progress in this direction has been already reported in \cite{Wen:2018agb,Fan:2020orx,Das:2021gts,Goto:2021sqx,Das:2022jrr} and more will be presented in the future work \cite{BGGN}.

\section*{Acknowledgement}
\addcontentsline{toc}{section}{Acknowledgement}
We are grateful to Alice Bernamonti, Adam Bzowski, Shira Chapman, Andrea Dei, Federico Galli, Sasha Gamayun, Felix Haehl, Matthew Headrick, Yasuaki Hikida, Romuald Janik, Monica Kang, Surbhi Khetrapal, Yuya Kusuki, Sinong Liu, Dimitrios Patramanis, Simon Ross, Bo Sundborg, Tadashi Takayanagi, Herman Verlinde, Xi Yin,  Claire Zukowski for discussions and comments. DG thanks the hospitality of GGI during the workshop ``Reconstructing the Gravitational Hologram with Quantum Information'' and the Jagiellonian University where part of the work was presented. PC would like to thank GGI and YITP Kyoto for hospitality during parts of this work. This work is supported by “Polish Returns 2019” grant of the National Agency for Academic Exchange (NAWA) PPN/PPO/2019/1/00010/U/0001 and Sonata Bis 9 2019/34/E/ST2/00123 grant from the National Science Centre, NCN.

\appendix
\section{$\mathfrak{sl}(2,\mathbb{R})$ Sectors of the Virasoro Algebra}

\subsection{Matrix representation of $\mathfrak{sl}
(2,\mathbb{R})$} \label{app:slmatrix}
The matrix representation for the generators of the special linear group $SL(2,\mathbb{R})$ can be adopted as the following,
\begin{align}
L_{-1} = 
\begin{pmatrix}
0& -1 \\
0&0
\end{pmatrix}\,,~~~
L_0 = \begin{pmatrix}
- {1\over 2}& 0 \\
0& {1\over 2}
\end{pmatrix}\,,~~~
L_1=\begin{pmatrix}
0& 0 \\
1&0
\end{pmatrix}\,.
\end{align}
It is easy to check that they satisfy the $\mathfrak{sl}(2,\mathbb{R})$ algebra $[L_m,L_n] = (m-n) L_{m-n}$ with $m,n\in \{0,\pm1\}$. We can then use this representation to check e.g. various BCH formulas (as below) etc.

\subsection{Closed form of BCH formula } \label{app:BCH}
Since the subalgebra formed by $L_0$ and $L_{\pm k}$ is essentially an $\mathfrak{sl}^{(k)}(2,\mathbb{R})$ algebra, it is expected that there should be some nice closed form of the BCH formula, as shown in \cite{Matone:2015wxa} for the Virasoro generators at level $k$
\begin{align}\label{eq:subsecVira}
\begin{array}{c}
\exp \left(\lambda_{-k} L_{-k}\right) \exp \left(\lambda_{0} L_{0}\right) \exp \left(\lambda_{k} L_{k}\right)= \\
\exp \left\{\frac{\lambda_{-}-\lambda_{+}}{e^{k \lambda_{-}}-e^{k \lambda_{+}}}\left[k \lambda_{-k} L_{-k}-\left(2-e^{k \lambda_{+}}-e^{k \lambda_{-}}\right) L_{0}+k \lambda_{k} L_{k}-c_{k} I\right]\right\}
\end{array}
\end{align}
where the coefficients $c_k$ and $\lambda_{\pm}$ are given by
\begin{align}
e^{k \lambda_{\pm}}&=\frac{1+e^{k \lambda_{0}}-k^{2} \lambda_{-k} \lambda_{k} \pm \sqrt{\left(1+e^{k \lambda_{0}}-k^{2} \lambda_{-k} \lambda_{k}\right)^{2}-4 e^{k \lambda_{0}}}}{2}\,,\\
c_{k}&=\frac{\lambda_{-k} \lambda_{k}}{\lambda_{+}-\lambda_{-}}\left(\frac{\lambda_{+}}{1-e^{k \lambda_{+}}}-\frac{\lambda_{-}}{1-e^{k \lambda_{-}}}\right) \frac{c}{12}\left(k^{4}-k^{2}\right)\,.
\end{align}
To kill the $L_0$ term in equation \eqref{eq:subsecVira}, we can set
\begin{align}
\lambda_+ &=   \frac{\ln(1+\sqrt{1-e^{k \lambda_0}})}{k} =  \frac{\ln (1+\tanh (k\rho))}{k}\,,\\\lambda_- &=    \frac{\ln(1-\sqrt{1-e^{k \lambda_0}})}{k} =   \frac{\ln (1-\tanh (k\rho))}{k}\,,
\end{align}
choosing $\lambda_0 = -\frac{2\ln (\cosh (k\rho))}{k}$. To make the operator in equation \eqref{eq:subsecVira} unitary, it is convenient to write $\lambda_{-k}$ and $\lambda_k$ in the following form,
\be
\lambda_k =- \frac{\tanh(k\rho)}{k}e^{-i\theta}\,,~~~ \lambda_{-k} =  \frac{\tanh(k\rho)}{k}e^{i\theta}
\ee 
substituting back into eq. \eqref{eq:subsecVira}, this gives the desired form
\begin{align}
&~~\exp \left(\frac{\tanh(k\rho)}{k}e^{i\theta}
 L_{-k}\right) \exp \left(- \frac{2\ln (\cosh(k\rho))}{k} (L_{0} + c(k^2-1)/24)\right) \exp \left(-\frac{\tanh(k\rho)}{k}e^{-i\theta} L_{k}\right)\no\\
 &=\exp \left( -\rho(e^{-i\theta}L_{k} -e^{i\theta}L_{-k}  ) \right)\,.
\end{align} 
\section{Expectation value of the stress tensor}\label{app:Texp}
This appendix contains a pedagogical and detailed derivation of the stress tensor expectation value. Even though the derivation uses nothing but standard tricks and results from the Virasoro algebra, we still present it for curious readers interested in following all the details of our derivation.
\subsection{Setup}
First, it will be convenient to split the (chiral) stress tensor operator as
\be
T(z)=\sum_{n\in\mathbb{Z}}z^{-n-2}L_n=z^{-2}\left(L_0+\sum^\infty_{n=1}(z^nL_{-n}+z^{-n}L_n)\right),
\ee
and write our coherent states as
\be
\ket{\Psi_k}=(1-z_k\bar{z}_k)^{h_k}\sum^\infty_{p=0}\frac{z^p_k}{p!k^p}L^p_{-k}\ket{h},\quad \bra{\Psi_k}=(1-z_k\bar{z}_k)^{h_k}\sum^\infty_{q=0}\frac{\bar{z}^q_k}{q!k^q}\bra{h}L^q_k,
\ee
and we used notation for coherent state parameters
\be
z_k=e^{i\theta}\tanh\left(k\rho\right),\qquad \bar{z}_k=e^{-i\theta}\tanh\left(k\rho\right).
\ee
In both formulas we used the Virasoro generators satisfying the algebra
\be
[L_n,L_m]=(n-m)L_{n+m}+\frac{c}{12}n(n^2-1)\delta_{n+m,0}.
\ee
Moreover, the generators used for our coherent states involve only a subset of three: $\{L_{-k},L_0,L_k\}$, for some fixed $k$ satisfying
\be
[L_{0},L_{\pm k}]=\mp kL_{\pm k},\qquad [L_k,L_{-k}]=2kL_0+\frac{c}{12}k(k^2-1),\label{VirAl}
\ee
that is an algebra isomorphic to $SL(2,R)$. Finally, the state $\ket{h}$ is the highest-weight state that satisfies
\be
L_n\ket{h}=0,\qquad \text{for}\quad n>0,\qquad \bra{h}L_n=0,\qquad \text{for}\quad n<0.
\ee
Our main goal is to compute the expectation value 
\be
z^2\bra{\Psi_k}T(z)\ket{\Psi_k}=\bra{\Psi_k}L_0\ket{\Psi_k}+\sum^\infty_{n=1}(z^n\bra{\Psi_k}L_{-n}\ket{\Psi_k}+z^{-n}\bra{\Psi_k}L_{n}\ket{\Psi_k}).
\ee
We will compute these terms separately, but first lets review some basic facts and tricks from Virasoro algebra that will be important to get the result.
\subsection{Basic tools and tricks}
The fact number one is that states with different eigenvalues of the generator $L_0$ are orthogonal. Since expectation values can be regarded as overlaps between different states, this will provide important constraints.\\
Next, we will accuire a few useful tools by performing simpler, intermediate computations. First, consider integers $n\ge 1$, $p>0$ and the state
\be
L_n L^p_{-k}\ket{h}=\left(L^p_{-k}L_n+\sum^{p-1}_{l=0}L^l_{-k}[L_n,L_{-k}]L^{p-1-l}_{-k}\right)\ket{h},
\ee
where the right hand side is simply a rewriting of the process of moving $L_n$ all the way to the right through $L^p_{-k}$'s. Next, we use the Virasoro algebra commutator
\be
[L_n,L_{-k}]=(n+k)L_{n-k}+\frac{c}{12}n(n^2-1)\delta_{n,k},
\ee
and the fact that $L_n$ annihilates $\ket{h}$, to write our first useful result
\be
L_n L^p_{-k}\ket{h}=\sum^{p-1}_{l=0}L^l_{-k}\left((n+k)L_{n-k}+\frac{c}{12}k(k^2-1)\delta_{n,k}\right)L^{p-1-l}_{-k}\ket{h}.\label{RES1}
\ee
For another stepping stone, we start from the $n=k$ case of the above formula
\bea
L_k L^p_{-k}\ket{h}&=&\sum^{p-1}_{l=0}L^l_{-k}\left(2kL_0+\frac{c}{12}k(k^2-1)\right)L^{p-1-l}_{-k}\ket{h},\nn\\
&=&\sum^{p-1}_{l=0}\left(2k(h+k(p-1-l))+\frac{c}{12}k(k^2-1)\right)L^{p-1}_{-k}\ket{h},\nn\\
&=&k^2p(2h_k+p-1)L^{p-1}_{-k}\ket{h}\equiv A^p_{k}\,L^{p-1}_{-k}\ket{h}.
\eea
In the second step we just used that the state on the right is an eigenstate of $L_0$ with eigenvalue $h+k(p-1-l)$ and then moved $L^l_{-k}$ through. Finally in the last line we defined
\be
A^p_{k}=k^2p(2h_k+p-1),\qquad h_k=\frac{1}{k}\left(h+\frac{c}{24}(k^2-1)\right).
\ee
This formula can be used recursively to compute the action of higher powers: $L^q_k$ given by
\be
L^q_k L^p_{-k}\ket{h}=L^{q-1}_k\left(L_kL^p_{-k}\ket{h}\right)=A^p_kL^{q-2}_k(L_kL^{p-1}_{-k}\ket{h})=A^p_kA^{p-1}_kL^{q-2}_kL^{p-2}_{-k}\ket{h},\label{LqLp}
\ee
and so on. From this observation we derive two facts: one that state \eqref{LqLp} vanishes for $q>p$ and second, the norm
\be
\bra{h}L^q_k L^p_{-k}\ket{h}=\delta^{q,p}\prod^p_{i=1}A^i_k=\delta^{q,p}p!k^{2p}\frac{\Gamma(2h_k+p)}{\Gamma(2h_k)}.\label{NORM}
\ee
For the following, it will be very useful to introduce a special notation for this norm
\be
\mathcal{N}_{k,p}\equiv p!k^{2p}\frac{\Gamma(2h_k+p)}{\Gamma(2h_k)},
\ee
in terms of which
\be
A^p_k=\frac{\mathcal{N}_{k,p}}{\mathcal{N}_{k,p-1}}.
\ee
Notice that in our recursive formula above these ratios cancel and we can write compactly
\be
L^l_k L^p_{-k}\ket{h}=\frac{\mathcal{N}_{k,p}}{\mathcal{N}_{k,p-1}}\frac{\mathcal{N}_{k,p-1}}{\mathcal{N}_{k,p-2}}...\frac{\mathcal{N}_{k,p-(l-1)}}{\mathcal{N}_{k,p-l}}L^{p-l}_{-k}\ket{h}=\frac{\mathcal{N}_{k,p}}{\mathcal{N}_{k,p-l}}L^{p-l}_{-k}\ket{h}.
\ee
Analogous result holds for the conjugate state
\be
\bra{h}L^{q}_k L^l_{-k}=\frac{\mathcal{N}_{k,q}}{\mathcal{N}_{k,q-l}}\bra{h}L^{q-l}_k,
\ee
and we will use it shortly. With these basic tools we are ready to compute the main matrix elements.
\subsection{$\bra{h}L^q_kL_{\pm n}L^p_{-k}\ket{h},$}
Lets now carefully compute the elements that will enter the expectation value. We start with $L_n$ and the building block 
\be
\bra{h}L^q_kL_{n}L^p_{-k}\ket{h}.
\ee
From our first fact that states with different eigenvalues of $L_0$ are orthogonal we must have
\be
kq+n=pq,
\ee
since $n$ is an integer, this correlator will be non-zero only for $n=k\tilde{n}$ as well as
\be
p=q+\tilde{n}.
\ee
We then need to compute (we drop the $\tilde{\,}$ for simple notation) :
\be
\bra{h}L^q_kL_{kn}L^{q+n}_{-k}\ket{h}.
\ee
Using our formula \eqref{RES1}, we can first write
\be
L_{kn} L^{q+n}_{-k}\ket{h}=\sum^{q+n-1}_{l=0}L^l_{-k}\left(k(n+1)L_{k(n-1)}+\frac{c}{12}k(k^2-1)\delta_{n,1}\right)L^{q+n-1-l}_{-k}\ket{h},
\ee
where we used $\delta_{kn,k}=\delta_{n,1}$. Then sandwiching it with $\bra{h}L^q_{k}$ gives
\bea
&&\bra{h}L^q_kL_{kn} L^{q+n}_{-k}\ket{h}=\sum^{q}_{l=0}\bra{h}L^q_kL^l_{-k}\left(k(n+1)L_{k(n-1)}+\frac{c}{12}k(k^2-1)\delta_{n,1}\right)L^{q+n-1-l}_{-k}\ket{h},\nn\\
&&=\sum^{q}_{l=0}\frac{\mathcal{N}_{k,q}}{\mathcal{N}_{k,q-l}}\left(k(n+1)\bra{h}L^{q-l}_kL_{k(n-1)}L^{q+n-1-l}_{-k}\ket{h}+\frac{c}{12}k(k^2-1)\delta_{n,1}\mathcal{N}_{k,q-l}\right),\nn\\
\eea
where in the first line we used that for $l>q$  we have $\bra{h}L^q_kL^l_{-k}=0$ and in the second line the norm \eqref{NORM}. We can also change the summation index to $l_1=q-l$ to write this as
\bea
\bra{h}L^q_kL_{kn} L^{q+n}_{-k}\ket{h}=\sum^{q}_{l_1=0}\frac{\mathcal{N}_{k,q}}{\mathcal{N}_{k,l_1}}\left(k(n+1)\bra{h}L^{l_1}_kL_{k(n-1)}L^{l_1+n-1}_{-k}\ket{h}+\frac{c}{12}k(k^2-1)\delta_{n,1}\mathcal{N}_{k,l_1}\right).\nn\\
\eea
This correlator will be very useful and we will put it into recursive form in a moment. First let us compute  the $n=1$ case
\bea
\bra{h}L^q_kL_{k} L^{q+1}_{-k}\ket{h}=2k^2\mathcal{N}_{k,q}\sum^{q}_{l_1=0}\left(h_k+l_1\right)=k^2(q+1)(2h_k+q)\mathcal{N}_{k,q}.\label{resn=1}
\eea
On the other hand, for $n>1$ we have
\bea
\bra{h}L^q_kL_{kn} L^{q+n}_{-k}\ket{h}=k(n+1)\sum^{q}_{l_1=0}\frac{\mathcal{N}_{k,q}}{\mathcal{N}_{k,l_1}}\bra{h}L^{l_1}_kL_{k(n-1)}L^{l_1+n-1}_{-k}\ket{h}.
\eea
Now we iterate this expression. The first step is
\bea
\bra{h}L^q_kL_{kn}L^{q+n}_{-k}|h\rangle=k^2(n+1)n\sum^{q}_{l_1=0}\frac{\mathcal{N}_{k,q}}{\mathcal{N}_{k,l_1}}\sum^{l_1}_{l_2=0}\frac{\mathcal{N}_{k,l_1}}{\mathcal{N}_{k,l_2}}\bra{h}L^{l_2}_kL_{k(n-2)}L^{l_2+n-2}_{-k}\ket{h}.
\eea
So basically we can just iterate this formula until we reach the correlator with $L_{k}$, that means we need to have $n-1$ sums:
\bea
\bra{h}L^q_kL_{kn}L^{q+n}_{-k}|h\rangle=k^{n-1}\left(\prod^{n-1}_{i=1}(n-i+2)\right)\sum^{q}_{l_1=0}...\sum^{l_{n-2}}_{l_{n-1}=0}\frac{\mathcal{N}_{k,q}}{\mathcal{N}_{k,l_{n-1}}}\bra{h}L^{l_{n-1}}_kL_{k}L^{l_{n-1}+1}_{-k}\ket{h}.\nn\\
\eea
Finally, we use our previous result \eqref{resn=1} but written as a sum
\bea
\bra{h}L^{l_{n-1}}_kL_{k}L^{l_{n-1}+1}_{-k}\ket{h}=2k^2\mathcal{N}_{k,l_{n-1}}\sum^{l_{n-1}}_{l_n=0}(h_k+l_n).
\eea
This way, our main expression becomes
\bea
\bra{h}L^q_kL_{kn}L^{q+n}_{-k}|h\rangle=k^{n+1}(n+1)!\mathcal{N}_{k,q}\sum^{q}_{l_1=0}...\sum^{l_{n-1}}_{l_{n}=0}(h_k+l_n).
\eea
Luckily, these nested sums can performed and compactly written as
\be
\sum^{q}_{l_1=0}...\sum^{l_{n-1}}_{l_{n}=0}(h_k+l_n)=\frac{(q+n)!}{q!(n+1)!}((n+1)h_k+q).
\ee
Finally, we can write our expression as
\bea
\bra{h}L^q_kL_{kn}L^{q+n}_{-k}|h\rangle&=&k^{n+1}\mathcal{N}_{k,q}\frac{(q+n)!}{q!}((n+1)h_k+q),\nn\\
&=&k^{2q+n+1}(q+n)!\frac{\Gamma(2h_k+q)}{\Gamma(2h_k)}((n+1)h_k+q),\label{MEqn}
\eea
and similarly for the conjugation
\be
\bra{h}L^{p+n}_kL_{-kn}L^p_{-k}\ket{h}=\left(\bra{h}L^p_{k}L_{kn}L^{p+n}_{-k}\ket{h}\right)^\dagger=k^{n+1}\mathcal{N}_{k,p}\frac{(p+n)!}{p!}((n+1)h_k+p).\label{MEpn}
\ee
Note also that both expressions agree with $n=\pm 1$ results.
\subsection{$\bra{\Psi_k}L_{\pm n}\ket{\Psi_k}$ and $\bra{\Psi_k}L_{0}\ket{\Psi_k}$}
Now we are ready to evaluate
\bea
\bra{\Psi_k}L_{kn}\ket{\Psi_k}&=&(1-z_k\bar{z}_k)^{2h_k}\sum^\infty_{q=0}\frac{z^{q+n}_k}{(q+n)!k^{q+n}}\frac{\bar{z}^q_k}{q!k^q}\bra{h}L^q_kL_{kn}L^{q+n}_{-k}\ket{h},\nn\\
&=&kz^n_k(1-z_k\bar{z}_k)^{2h_k}\sum^\infty_{q=0}\frac{(z_k\bar{z}_z)^q}{q!}\frac{\Gamma(2h_k+q)}{\Gamma(2h_k)}((n+1)h_k+q),\nn\\
&=&kh_k z^n_k\left(n-1+\frac{2}{1-z_k\bar{z}_k}\right),
\eea
where we used the result \eqref{MEqn}. Similarly, with \eqref{MEpn}, we carefully  get
\bea
\bra{\Psi_k}L_{-kn}\ket{\Psi_k}&=&(1-z_k\bar{z}_k)^{2h_k}\bar{z}^n_k\sum^\infty_{p=0}\frac{(z_k\bar{z}_k)^{p}}{p!(p+n)!k^{2p+n}}\bra{h}L^{p+n}_kL_{-kn}L^p_{-k}\ket{h},\nn\\
&=&k\bar{z}^n_k(1-z_k\bar{z}_k)^{2h_k}\sum^\infty_{p=0}\frac{(z_k\bar{z}_k)^{p}}{p!}\frac{\Gamma(2h_k+p)}{\Gamma(2h_k)}((n+1)h_k+p),\nn\\
&=&kh_k \bar{z}^n_k\left(n-1+\frac{2}{1-z_k\bar{z}_k}\right),
\eea
that is of course a complex conjugate of the previous element.
Las but not the least, we compute the $L_0$ element
\bea
\bra{\Psi_k}L_0\ket{\Psi_k}&=&(1-z_k\bar{z}_k)^{2h_k}\sum^\infty_{p,q=0}\frac{z^p_k}{p!k^p}\frac{\bar{z}^q_k}{q!k^q}\bra{h}L^q_kL_0L^p_{-k}\ket{h},\nn\\
&=&(1-z_k\bar{z}_k)^{2h_k}\sum^\infty_{p,q=0}\frac{z^p_k}{p!k^p}\frac{\bar{z}^q_k}{q!k^q}(h+kp)\bra{h}L^q_kL^p_{-k}\ket{h},\nn\\
&=&(1-z_k\bar{z}_k)^{2h_k}\sum^\infty_{p=0}\frac{z^p_k\bar{z}^p_k}{p!}(h+kp)\frac{\Gamma(2h_k+p)}{\Gamma(2h_k)},\nn\\
&=&h+\frac{2h_kk z_k\bar{z}_k}{1-z_k\bar{z}_k}.
\eea
\subsection{Re-summation}
Finally we can sum both contributions
\be
\sum^{\infty}_{n=1}\bra{\Psi_k}z^{kn}L_{-kn}+z^{-kn}L_{kn}\ket{\Psi_k}=kh_k\sum^{\infty}_{n=1}\left(n-1+\frac{2}{1-z_k\bar{z}_k}\right)\left(\left(z^{k}\bar{z}_k\right)^n+\left(z^{-k}z_k\right)^n\right),
\ee
which sums up to
\be
kh_k\left(\frac{z^2_k}{(z^k-z_k)^2}+\frac{2z_k}{(z^k-z_k)(1-z_k\bar{z}_k)}\right)+kh_k\left(\frac{\bar{z}^2_k}{(z^{-k}-
\bar{z}_k)^2}+\frac{2\bar{z}_k}{(z^{-k}-\bar{z}_k)(1-z_k\bar{z}_k)}\right).
\ee 
Putting everything together
\be
z^2\bra{\Psi_k}T(z)\ket{\Psi_k}=\bra{\Psi_k}L_0\ket{\Psi_k}+\sum^\infty_{n=1}(z^{kn}\bra{\Psi_k}L_{-kn}\ket{\Psi_k}+z^{-kn}\bra{\Psi_k}L_{kn}\ket{\Psi_k}),
\ee
allows us to write
\be
\bra{\Psi_k}T(z)\ket{\Psi_k}=\frac{kh_kz^{2(k-1)}(1-z_k\bar{z}_k)^2}{(z^k-z_k)^2(1-z^k\bar{z}_k)^2}-\frac{c}{24}(k^2-1)\frac{1}{z^2}.
\ee
This is our result \eqref{eq:SEexp}.

\section{Geodesics in AdS from the embedding formalism} 
In this section, we obtain the most general form of the trajectory of a massive particle using the embedding formalism  both in Lorentzian and Euclidean signatures. We start in Lorentzian signature, where the trajectory is essentially a timelike geodesics. 
The global Lorentzian AdS$_3$ can be embedded in $\mathbb{R}^{2,2}$ in terms of the coordinates
\begin{align}\label{eq:geodesics}
X_0 &= \sec\phi \cos t\,,~~~ X_{0'} = \sec \phi \sin t\,,\\
X_1 &= \tan\phi \cos \theta\,,~~~ X_2 = \tan\phi \sin \theta\,.
\end{align}
The  the trajectory of the particle in the embedding coordinates  has to satisfy the constraining equation \cite{Dorn:2005jt}
\be
E X_\mu = J_{0\mu}X_{0'} - J_{0'\mu}X_0
\ee
where $J_{0\mu}$ and $J_{0'\mu}$ are the conserved charges. Redefining them as
\be
\zeta_\mu \equiv  J_{0'\mu} - i J_{0\mu}\,,~~~ \zeta_\mu^* \equiv  J_{0'\mu} + i J_{0\mu}
\ee
we can be expressed the trajectory of the massive particle in terms of the embedding coordinates as 
\begin{align}
X_0 &= \sec\phi (t) \cos t\,,~~~ X_{0'} = \sec \phi (t) \sin t\,,\\
X_\mu (t) &= - \frac{1}{2E\cos\phi(t)} (\zeta_\mu e^{-i t} + \zeta_\mu^* e^{it})\,,
\end{align}
to stay on the hyperboloid, one has
\be
\sin\phi(t) = \frac{1}{2E}\sqrt{e^{-2it} \zeta^2 + 2\zeta \cdot \zeta^* +e^{2it} \zeta^{*2}}\,.
\ee
Comparing with eq. \eqref{eq:geodesics}, 
the geodesic equations parametrized by $t$ are given as
\begin{align}
\sin\phi &= \frac{1}{2E}\sqrt{e^{-2it} \zeta^2 + 2\zeta \cdot \zeta^* +e^{2it} \zeta^{*2}}\,,\\
\tan\theta &= \frac{\zeta^*_1e^{it} + \zeta_1 e^{-it}}{\zeta^*_2 e^{it} +\zeta_2 e^{-it}}\,,
\end{align}
actually in the first equation, the energy $E$ can be absorbed into the  parameters, while it doesn't affect the second equation. Let us denote the new parameters to be $\tilde \zeta_\mu = \zeta_\mu/(2E)$, to comply with the range of the coordinates $\phi$ and $\theta$, we have $\tilde \zeta\cdot \tilde \zeta^*<1$. In total, there are four real parameters, which is reasonable, since we only need to specify the position and velocity of the particle on a given Cauchy slice for AdS$_3$.

\paragraph{Euclidean case}
The global Euclidean AdS$_3$ can be embedded in $\mathbb{R}^{1,3}$ in terms of the coordinates
\begin{align}
X_0 &= \cosh \rho \cosh \tau\,,~~~ X_{0'} = \cosh \rho \sinh \tau\,,\\
X_1 &= \sinh \rho \cos \theta\,,~~~ X_2 = \sinh \rho \sin \theta\,.
\end{align}
using a change of variable $\tanh\rho = \sin\phi$, one has the embedding
\begin{align}\label{eq:embeddingGL}
X_0 &=\sec\phi\cosh \tau\,,~~~ X_{0'} = \sec\phi \sinh \tau\,,\\
X_1 &= \tan\phi \cos \theta\,,~~~ X_2 = \tan\phi \sin \theta\,.
\end{align}
As a comparison to the Poincar\'e embedding and other global embedding, we can write those relations as
\begin{align}
X_0 &= \frac{l^2 + z^2 +x^2+t^2}{2z} = \sqrt{l^2 +r^2}\cosh\tau = \frac{l\cosh\tau}{\sin\phi}\,,\no\\
X_1 &= \frac{l t}{z} =  \sqrt{l^2 +r^2}\sinh\tau =\frac{l\sinh\tau}{\sin\phi} \,,\no\\
 X_2 &= \frac{l x}{z} = r \sin\theta = l\tan\phi \sin\theta\,, \no\\
 X_3 &= \frac{-l^2 + z^2 +x^2+t^2}{2z} = r\cos\theta= l\tan\phi \cos\theta\,.
 \end{align}

In Euclidean case, the geodesics are obtained by taking the solutions $t\to i\tau$ in
 \eqref{eq:geodesics}, and change $\zeta_\mu \to \alpha_\mu$ and $\zeta^*_\mu\to \beta_\mu$ as independent parameters. The geodesics parametrized in terms of $\tau$ are given as
 \begin{align}
 \sin\phi &= \frac{1}{2E}\sqrt{e^{2\tau} \alpha^2 + 2\alpha \cdot \beta +e^{-2\tau} \beta^{2}}\,,\\
\tan\theta &= \frac{\beta_1e^{-\tau} + \alpha_1 e^{\tau}}{\beta_2 e^{-\tau} +\alpha_2 e^{\tau}}\,,
 \end{align}
in terms of the embedding coordinates
\begin{align}
X_0 &= \sec\phi (\tau) \cosh \tau\,,~~~ X_{0'} = \sec \phi (\tau) \sinh\tau\,,\\
X_\mu (\tau) &= - \frac{1}{2E\cos\phi(\tau)} (\alpha_\mu e^{\tau} + \beta_\mu e^{-\tau})\,.
\end{align}
We are interested in one particular type of the geodesics passing through $(\tau = \pm \tau_0,\phi = \pi/2)$ on the constant $\theta=\theta_0$ plane, such boundary conditions lead to $\alpha_1=\alpha_2\tan\theta_0$ and $\beta_1 = \beta_2\tan\theta_0$, 
taking $\alpha_2 =\beta_2= \frac{\cos\theta_0}{2\cosh\tau_0}$, we have
\be\label{eq:globalgeo}
\sin\phi =\frac{\cosh\tau}{\cosh\tau_0}\,,~~~\theta=\theta_0\,,
\ee
which reproduces eq. \eqref{eq:geodesictau0} in the main text.

\bibliographystyle{utphys.bst}
\bibliography{bibEE}{}

\end{document}